\documentclass[aps,twocolumn,superscriptaddress,footinbib,floatfix,showpacs]{revtex4}

\usepackage{graphicx}
\usepackage{amsmath}

\def\ket#1{|{#1}\rangle}
\def\bra#1{\langle{#1}|}
\newcommand{\dlangle}{\left\langle\!\left\langle}
\newcommand{\drangle}{\right\rangle\!\right\rangle}
\def\dexpct#1{\dlangle{#1}\drangle}
\def\expct#1{\!\left\langle{#1}\right\rangle}
\def\eket{\ket{\mathrm{e}}}
\def\gket{\ket{\mathrm{g}}}
\def\ebra{\bra{\mathrm{e}}}
\def\gbra{\bra{\mathrm{g}}}
\def\goneket{\ket{\mathrm{g}, 1}}

\def\HAF{H_\mathrm{\scriptscriptstyle AF}}
\def\HA{H_\mathrm{\scriptscriptstyle A}}
\def\HCM{H_\mathrm{\scriptscriptstyle CM}}
\def\HD{H_\mathrm{\scriptscriptstyle D}}
\def\DSE{D_\mathrm{\scriptscriptstyle SE}}

\def\kL{k_\mathrm{\scriptscriptstyle L}}
\def\kLbf{\mathbf{k}_\mathrm{\scriptscriptstyle L}}

\def\psiket{{| \psi \rangle}}
\def\ce{c_\mathrm{e}}
\def\cg{c_\mathrm{g}}
\def\ckz{c_{\mathbf{k},\zeta}}
\def\cet{\tilde c_\mathrm{e}}
\def\cgt{\tilde c_\mathrm{g}}
\def\ckzt{\tilde c_{\mathbf{k},\zeta}}
\def\gkzket{{| \mathrm{g}, 1_{\mathrm{k},\zeta} \rangle}}
\def\omegak{\omega_\mathbf{k}}

\def\ss{_{\mathrm{ss}}}
\def\rhoext{\rho_\mathrm{ext}}
\def\etaeff{\eta_\mathrm{eff}}
\def\keff{k_\mathrm{eff}}

\def\eqnarr#1#2{  
\renewcommand{\arraystretch}{#1}
  \setlength\arraycolsep{0ex}
  \begin{array}{rcl}
    #2
  \end{array}
}
\def\ds{\displaystyle}
\def\arreq{&{}={}&\ds }

\def\Vx{V_X}
\def\Vp{V_P}
\def\Cxp{C_{XP}}

\begin{document}

\title{A Straightforward Introduction to Continuous Quantum Measurement}

\author{Kurt Jacobs}
\affiliation{Department of Physics, University of Massachusetts at Boston, Boston, MA 02124}
\affiliation{Quantum Sciences and Technologies Group, Hearne Institute for Theoretical Physics, Department of Physics \& Astronomy, Louisiana State University, Baton Rouge, LA 70803-4001}

\author{Daniel A. Steck}
\affiliation{Department of Physics and Oregon Center for Optics, 1274 University of Oregon, Eugene, OR 97403-1274}

\begin{abstract}
We present a pedagogical treatment of the formalism of continuous
quantum measurement.  Our aim is to show the reader how the equations
describing such measurements are derived and manipulated in a direct
manner.  We also give elementary background material for those new to
measurement theory, and describe further various aspects of continuous
measurements that should be helpful to those wanting to use such
measurements in applications.  Specifically, we use the simple
and direct approach of
generalized measurements to derive the stochastic master equation
describing the continuous measurements of observables, give a tutorial
on stochastic calculus, treat multiple observers and inefficient
detection, examine a general form of the measurement master equation,
and show how the master equation leads to information gain and
disturbance.  To conclude, we give a detailed treatment of imaging the
resonance fluorescence from a single atom as a concrete example of how
a continuous position measurement arises in a physical system.
\end{abstract}

\maketitle

\section{Introduction}

When measurement is first introduced to students of quantum mechanics,
it is invariably treated by ignoring any consideration of the time the measurement takes:
the measurement just ``happens,'' for all intents and purposes,
instantaneously.  This treatment is good for a first introduction, but
is not sufficient to describe two important situations.  The first is
when some aspect of a system is continually monitored.  
This happens, for example,
when one illuminates an object and continually detects the reflected
light in order to track the object's motion.  In this case, information
is obtained about the object at a finite rate, and one needs to
understand what happens to the object {\em while} the measurement
takes place.  It is the subject of {\em continuous quantum
measurement} that describes such a measurement.  The second situation
arises because nothing really happens instantaneously.  Even
rapid, ``single shot'' measurements take some time.  If this time is
not short compared to the dynamics of the measured system, then it is
once again important to understand both the dynamics of the flow of
information to the observer and the effect of the measurement on the
system.

Continuous measurement has become increasingly important in the last
decade, due mainly to the growing interest in the application of
\textit{feedback control} in quantum
systems~\cite{BelavkinLQG,DJ,Wiseman05,Hopkins03,Steck04,Steixner05,Rabl05,Combes06,Bushev06,DHelon06,Steck06}.  
In feedback control a system is
continuously measured, and this information is used while the
measurement proceeds (that is, in real time) to modify the system
Hamiltonian so as to obtain some desired behavior.  Thus, continuous
measurement theory is essential for describing feedback control.  The
increasing interest in continuous measurement is also due to its
applications in metrology~\cite{Wiseman95,Berry02,Pope04,Stockton04,Geremia05},
quantum information~\cite{Dolinar73,Geremia04,Jacobs07}, quantum
computing~\cite{Ahn02,Sarovar04,Handel06}, and its importance in
understanding the quantum to classical
transition~\cite{Bhattacharya00,Habib02,Bhattacharya03,Ghose04,Ghose05,Everitt05,Habib06}.

While the importance of continuous measurement grows, to date there is 
really only one introduction to the subject that could be described as 
both easily accessible and extensive, that being the one by Brun in the 
American Journal of Physics~\cite{Brun02} (some other pedagogical treatments 
can be found in~\cite{Braginsky95,Carm93,WisemanLinQ}).  While the
analysis in Brun's work is suitably direct, it treats explicitly only
measurements on two-state systems, and due to their simplicity the
derivations used there do not easily extend to measurements of more
general observables.  Since many applications involve measurements of
observables in infinite-dimensional systems (such as the position of a
particle), we felt that an introductory article that derived the
equations for such measurements in the simplest and most direct
fashion would fill an important gap in the literature.  This is what we
do here.  Don't be put off by the length of this article---a reading
of only a fraction of the article is sufficient to understand how to
derive the basic equation that describes continuous measurement, the
mathematics required to manipulate it (the so-called It\^o calculus),
and how it can be solved.  This is achieved in Sections
\ref{section:continuousmeasurement}, \ref{section:Ito}, and
\ref{section:solution}.  If the reader is not familiar with the
density operator, then this preliminary material is explained in
Section~\ref{section:basic}, and generalized quantum measurements
(POVM's) are explained in Section~\ref{section:weak}.

The rest of the article gives some more information about continuous
measurements.  In Section~\ref{section:multipleobservers} we show how
to treat multiple, simultaneous observers and inefficient detectors,
both of which involve simple and quite straightforward generalizations
of the basic equation.  In Section~\ref{section:genform} we discuss
the most general form that the continuous-measurement equation can
take.  In Section~\ref{section:interpretation} we present explicit
calculations to explain the meaning of the various terms in the
measurement equation.  Since our goal in the first part of this
article was to derive a continuous measurement equation in the
shortest and most direct manner, this did not involve a concrete
physical example.  In the second-to-last (and longest) section, we
provide such an example, showing in considerable detail how a
continuous measurement arises when the position of an atom is
monitored by detecting the photons it emits.  The final section
concludes with some pointers for further reading.

\section{Describing an Observer's State of Knowledge of a Quantum System}
\label{section:basic}

\subsection{The Density Operator}

Before getting on with measurements, we will briefly review the density
operator, since it is so central to our discussion.
The density operator represents the state of a quantum system
in a more general way than the state vector, and equivalently
represents an observer's \textit{state of knowledge} of a system.

When a quantum state can be represented by a state vector $\ket\psi$,
the density operator is defined as the product
\begin{equation}
  \rho := \ket\psi\bra\psi.
  \label{rhopure}
\end{equation}
In this case, it is obvious that the information content
of the density operator is equivalent to that of the state vector
(except for the overall phase, which is not of physical significance).

The state vector can represent states of \textit{coherent} superposition.
The power of the density operator lies in the fact that it can 
represent \textit{incoherent} superpositions as well.
For example, let $\ket{\psi_\alpha}$ be a set of states (without any 
particular restrictions).  Then the density operator 
\begin{equation}
  \rho = \sum_\alpha p_\alpha \ket{\psi_\alpha}\bra{\psi_\alpha}
  \label{rhogen}
\end{equation}
models the fact that we don't know \textit{which} of the
states $\ket{\psi_\alpha}$ the system is in, but we know that
it is in the state $\ket{\psi_\alpha}$ with probability $p_\alpha$.
Another way to say it is this:
the state vector $\ket\psi$ represents a certain
\textit{intrinsic uncertainty} with respect to quantum observables;
the density operator can represent uncertainty \textit{beyond} the minimum
required by quantum mechanics.
Equivalently, the density operator can represent an \textit{ensemble}
of identical systems in possibly different states.

A state of the form $(\ref{rhopure})$ is said to be a \textit{pure state}.
One that cannot be written in this form is said
to be \textit{mixed}, and can be written in the form (\ref{rhogen}).

Differentiating the density operator and employing the
Schr\"odinger equation $i\hbar \partial_t \ket\psi = H\ket\psi$,
we can write down the equation of motion for the density operator:
\begin{equation}
  \partial_t\rho = -\,\frac{i}{\hbar}[H,\rho].
  \label{schrodvonneuman}
\end{equation}
This is referred to as the \textit{Schr\"odinger--von Neumann equation}.
Of course, the use of the density operator allows us to write down
more general evolution equations than those implied by 
state-vector dynamics.

\subsection{Expectation Values}

We can compute expectation values with respect to the density operator
via the trace operation. The trace of an operator $A$ is simply the
sum over the diagonal matrix elements with respect to any
complete, orthonormal set of states $\ket\beta$:
\begin{equation}
  \mathrm{Tr}[A]:= \sum_\beta\bra\beta  A\ket\beta .
\end{equation}
An important property of the trace is that the trace of a product is
invariant under cyclic permutations of the product.  For example,
for three operators,
\begin{equation}
  \mathrm{Tr}[ABC]
  =\mathrm{Tr}[BCA]
  =\mathrm{Tr}[CAB].
  \label{traceperm}
\end{equation}
This amounts to simply an interchange in the order of summations. 
For example, for two operators, working in the position representation,
we can use the fact that $\int dx\,\langle x\ket{x}$ is the identity operator
to see that
\begin{equation}
  \eqnarr{1.5}{
  \mathrm{Tr}[AB] \arreq \int dx\bra{x}AB\ket{x}\\
                  \arreq \int dx \int dx'\, \bra{x}A\ket{x'}\bra{x'}B\ket{x}\\
                  \arreq \int dx' \int dx \, \bra{x'}B\ket{x}\bra{x}A\ket{x'}\\
                  \arreq \int dx' \bra{x'}BA\ket{x'}\\
                  \arreq \mathrm{Tr}[BA].
  }
\end{equation}
Note that this argument assumes sufficiently ``nice'' operators
(it fails, for example, for $\mathrm{Tr}[xp]$).
More general permutations [e.g., of the form (\ref{traceperm})]
are obtained by replacements of the form $B\longrightarrow BC$.
Using this property, we can write the expectation
value with respect to a pure state as
\begin{equation}
  \expct{A} = \expct{\psi|A|\psi} = \mathrm{Tr}[A\rho].
\end{equation}
This argument extends to the more general
form (\ref{rhogen}) of the density operator.

\subsection{The Density Matrix}

The physical content of the density matrix is more apparent when 
we compute the elements $\rho_{\alpha\alpha'}$
of the \textit{density matrix} with respect to a complete, orthonormal basis.
The density matrix elements are given by
\begin{equation}
  \rho_{\alpha\alpha'} := \bra{\alpha}\rho\ket{\alpha'}.
\end{equation}
To analyze these matrix elements, we will assume the simple form
$\rho = \ket\psi\bra\psi$ of the density operator, though the
arguments generalize easily to arbitrary density operators.

The diagonal elements $\rho_{\alpha\alpha}$
are referred to as \textit{populations}, and give the probability
of being in the state $\ket\alpha$:
\begin{equation}
  \rho_{\alpha\alpha}=\bra{\alpha}\rho\ket{\alpha}
   = \left|\expct{\alpha|\psi}\right|^2.
\end{equation}
The off-diagonal elements $\rho_{\alpha\alpha'}$ (with $\alpha\neq\alpha'$)
are referred to as \textit{coherences}, since they give
information about the relative
phase of different components of the superposition.
For example, if we write the state vector as a superposition with
explicit phases,
\begin{equation}
  \ket\psi = \sum_\alpha c_\alpha \ket\alpha = \sum_\alpha |c_\alpha| e^{i\phi_\alpha}\ket\alpha,
\end{equation}
then the coherences are
\begin{equation}
  \rho_{\alpha\alpha'} = |c_\alpha c_{\alpha'}| e^{i(\phi_\alpha-\phi_{\alpha'})}.
\end{equation}
Notice that for a density operator not corresponding to a pure
state, the coherences in general will be the sum of complex numbers
corresponding to different states in the incoherent sum.
The phases will not in general line up, so that 
while $|\rho_{\alpha\alpha'}|^2 = \rho_{\alpha\alpha}\rho_{\alpha'\alpha'}$
for a pure state,
we expect $|\rho_{\alpha\alpha'}|^2 < \rho_{\alpha\alpha}\rho_{\alpha'\alpha'}$ 
($\alpha\neq\alpha'$) for a generic mixed
state.

\subsection{Purity}

The difference between pure and mixed states can be formalized in
another way.  Notice that the diagonal elements of the density matrix
form a probability distribution.  Proper normalization thus requires
\begin{equation}
  \mathrm{Tr}[\rho] = \sum_\alpha \rho_{\alpha\alpha} = 1.
\end{equation}
We can do the same computation for $\rho^2$, and we will
define the \textit{purity} to be $\mathrm{Tr}[\rho^2]$.
For a pure state,
the purity is simple to calculate:
\begin{equation}
  \mathrm{Tr}[\rho^2] = \mathrm{Tr}[\ket\psi\langle\psi|\psi\rangle\bra\psi]
  = \mathrm{Tr}[\rho] = 1.
\end{equation}
But for mixed states, $\mathrm{Tr}[\rho^2] < 1$.  For example, 
for the density operator in (\ref{rhogen}),
\begin{equation}
  \mathrm{Tr}[\rho^2] = \sum_\alpha p_\alpha^{\,2},
\end{equation}
if we assume the states $\ket{\psi_\alpha}$ to be orthonormal.
For equal probability of being in $N$ such states, 
$\mathrm{Tr}[\rho^2]=1/N$.  Intuitively, then, we can see
that $\mathrm{Tr}[\rho^2]$ drops to zero as the state becomes
more mixed---that is, as it becomes an incoherent superposition
of more and more orthogonal states.

\section{Weak Measurements and POVM's}
\label{section:weak}

In undergraduate courses the only kind of measurement that is
usually discussed is one in which the system is projected onto one
of the possible eigenstates of a given observable. If we write
these eigenstates as $\{|n\rangle :
n=1,\ldots,n_\mathrm{max}\}$, 
and the state of the
system is $|\psi\rangle = \sum_n c_n |n\rangle$, the probability
that the system is projected onto $|n\rangle$ is $|c_n|^2$. In
fact, these kind of measurements, which are often referred to as
{\em von Neumann} measurements, represent only a special class of
all the possible measurements that can be made on quantum
systems. However, all measurements can be derived from von Neumann
measurements.

One reason that we need to consider a larger class of measurements
is so we can describe measurements that extract only partial
information about an observable. A von Neumann measurement
provides complete information---after the measurement is
performed we know exactly what the value of the observable is,
since the system is projected into an eigenstate. Naturally,
however, there exist many measurements which, while reducing on
average our uncertainty regarding the observable of interest, do
not remove it completely.

First, it is worth noting that a von Neumann measurement can be
described by using a set of projection operators $\{P_n = |n\rangle\langle n
|\}$. Each of these operators describes what happens on one of the
possible outcomes of the measurement: 
if the initial state of the system is $\rho
= |\psi\rangle\langle\psi| $, 
then the $n$th possible outcome of the final state is 
given by
\begin{equation}
   \rho_\mathrm{f} 
     = |n\rangle\langle n|
     = \frac{P_n \rho P_n}{\mathrm{Tr}[P_n \rho P_n]},
\end{equation}
and this result is obtained with probability
\begin{equation}
  P(n) = \mbox{Tr}[P_n\rho P_n] = c_n ,
\end{equation}
where $c_n$ defines the superposition of the initial state
$\ket\psi$ given above.
It turns out that \textit{every} possible measurement may be described in a
similar fashion by generalizing the set of operators. 
Suppose we pick a set of $m_\mathrm{max}$ 
operators $\Omega_m$, the
only restriction being that $\sum_{m=1}^{m_\mathrm{max}} \Omega_m^\dagger\Omega_m
= I$, where $I$ is the identity operator. 
Then it is in principle possible to design a measurement
that has $N$ possible outcomes, 
\begin{equation}
   \rho_\mathrm{f} = \frac{\Omega_m \rho \Omega_m^\dagger }
        {\mathrm{Tr}[\Omega_m \rho \Omega_m^\dagger ]},
\end{equation}
with
\begin{equation}
 P(m) = \mathrm{Tr}[\Omega_m \rho \Omega_m^\dagger ]
\end{equation}
giving the probability of obtaining the $m$th outcome.

Every one of these more general measurements may be implemented by
performing a unitary interaction between the system and an
auxiliary system, and then performing a von Neumann measurement on
the auxiliary system. Thus all possible measurements may be
derived from the basic postulates of unitary evolution and von 
Neumann measurement~\cite{Schumacher96,mikeandike}. 

These ``generalized'' measurements are often referred to as  
POVM's, where the acronym stands for ``positive operator-valued measure.'' 
The reason for this is somewhat technical, but we explain it here because 
the terminology is so common. Note that the probability for obtaining 
a result in the range $[a,b]$ is 
\begin{equation}
  P(m\in [a,b]) = \sum_{m=a}^{b}\mbox{Tr}\left[\Omega_m\rho  \Omega_m^\dagger\right] = \mbox{Tr}\left[\sum_{m=a}^{b} \Omega_m^\dagger\Omega_m\rho\right]  .
\end{equation}
The positive operator $M = \sum_{m=a}^{b}\Omega_m^\dagger\Omega_m$ thus 
determines the probability that $m$ lies in the subset $[a,b]$ of its range. In this way 
the formalism associates a positive operator with {\em every} subset of the 
range of $m$, and is therefore a {\em positive operator-valued measure}.

Let us now put this into practice to describe a measurement that
provides partial information about an observable. In this case,
instead of our measurement operators $\Omega_m$ being projectors
onto a single eigenstate, we choose them to be a weighted sum of
projectors onto the eigenstates $|n\rangle$, each one peaked about
a different value of the observable. Let us assume now, for the
sake of simplicity, that the eigenvalues $n$ of the observable
$N$ 
take on all the integer values. In this case we can choose
\begin{equation}
   \Omega_m = \frac{1}{\cal N} \sum_{n} e^{-k(n-m)^2/4} |n\rangle\langle
   n| ,
\end{equation}
where ${\cal N}$ is a normalization constant chosen so
that $\sum_{m=-\infty}^\infty \Omega_m^\dagger\Omega_m = I$. We have now
constructed a measurement that provides partial information about
the observable $N$. This is illustrated clearly by examining the
case where we start with no information about the system. In this case 
the density matrix is completely mixed, so that $\rho \propto I$.
After making the measurement and obtaining the result $m$, the
state of the system is
\begin{equation}
  \rho_{\mbox{\scriptsize f}} = 
   \frac{\Omega_m \rho \Omega_m^\dagger }{\mbox{Tr}[\Omega_m \rho \Omega_m^\dagger 
   ]} = \frac{1}{\cal N} \sum_{n} e^{-k(n-m)^2/2} |n\rangle\langle
   n| .
\end{equation}
The final state is thus peaked about the eigenvalue $m$, but has a
width given by $1/\sqrt{k}$. The larger $k$, the less our final
uncertainty regarding the value of the observable. Measurements
for which $k$ is large are often referred to as {\em strong}
measurements, and conversely those for which $k$ is small are {\em
weak} measurements~\cite{FJ}. These are the kinds of measurements that we
will need in order to derive a continuous measurement in the next
section.

\section{A Continuous Measurement of an Observable}
\label{section:continuousmeasurement}

A continuous measurement is one in which information is
continually extracted from a system. Another way to say this is
that when one is making such a measurement, the amount of
information obtained goes to zero as the duration of the
measurement goes to zero. To construct a measurement like this, we
can divide time into a sequence of intervals of length $\Delta t$,
and consider a weak measurement in each interval. To obtain a
continuous measurement, we make the strength of each measurement
proportional to the time interval, and then take the limit in
which the time intervals become infinitesimally short.

In what follows, we will denote the observable we are measuring by
$X$ (i.e., $X$ is a Hermitian operator), and 
we will assume that it has a continuous spectrum of eigenvalues
$x$. We will write the eigenstates as $|x\rangle$, so that
$\langle x|x'\rangle = \delta(x-x')$. However, the equation that
we will derive will be valid for measurements of \textit{any} Hermitian
operator.

We now divide time into intervals of length $\Delta t$. In each
time interval, we will make a measurement described by the
operators
\begin{equation}
A(\alpha) = \left( \frac{4k\Delta t}{\pi} \right)^{1/4} \int_{-\infty}^{\infty}
e^{-2k\Delta t(x-\alpha)^2} |x\rangle\langle x| dx .
  \label{xpovm}
\end{equation}
Each operator $A(\alpha)$ a Gaussian-weighted sum of projectors  
onto the eigenstates of $X$. 
Here $\alpha$ is a continuous 
index, so that there is a continuum of measurement results labeled by $\alpha$.

The first thing we need to know is the probability density $P(\alpha)$ of the
measurement result $\alpha$ when $\Delta t$ is small. To work this 
out we first calculate the mean value of $\alpha$. If the initial state is 
$|\psi\rangle = \int \psi(x)|x\rangle dx$ then
$P(\alpha) =  \mbox{Tr}[A(\alpha)^\dagger A(\alpha) |\psi\rangle\langle\psi |]$, and we 
have 
\begin{equation}
  \eqnarr{2.3}{
  \langle\alpha\rangle \arreq \int_{-\infty}^{\infty} \alpha P(\alpha) \, d\alpha \\
  \arreq \int_{-\infty}^{\infty} \alpha \mbox{Tr}[A(\alpha)^\dagger A(\alpha) |\psi\rangle\langle\psi |] \, d\alpha \\
                     \arreq \sqrt{\frac{4k\Delta t}{\pi}} \int_{-\infty}^{\infty} \int_{-\infty}^{\infty} \alpha |\psi(x)|^2 e^{-4k\Delta t(x-\alpha)^2} \, dx\, d\alpha  \\
                     \arreq \int_{-\infty}^{\infty} x |\psi(x)|^2 \, dx 
                     =\langle X \rangle .
  }
\end{equation}
To obtain $P(\alpha)$ we now write
\begin{equation}
  \eqnarr{2.2}{
P(\alpha) \arreq \mbox{Tr}[A(\alpha)^\dagger A(\alpha) |\psi\rangle\langle\psi |]  \\
          \arreq \sqrt{\frac{4k\Delta t}{\pi}} \int_{-\infty}^{\infty} |\psi(x)|^2   e^{-4k\Delta t(x-\alpha)^2} dx .
  }
  \label{probalphaPOVM}
\end{equation}
If $\Delta t$ is sufficiently small then the Gaussian is much
broader than $\psi(x)$. This means we can approximate $|\psi(x)|^2$ 
by a delta function, which must be centered at 
the expected position $\langle X\rangle$ so 
that $\langle\alpha\rangle = \langle X\rangle$ as calculated above. 
We therefore have 
\begin{equation}
  \eqnarr{2.2}{
P(\alpha) &{}\approx{}&\ds \sqrt{\frac{4k\Delta t}{\pi}} \int_{-\infty}^{\infty} \delta(x-\expct{X})   e^{-4k\Delta t(x-\alpha)^2} \; dx \\
          \arreq \sqrt{\frac{4k\Delta t}{\pi}} e^{-4k\Delta t(\alpha - \langle X\rangle)^2} .
  }
\end{equation}
We can also write $\alpha$ as the stochastic quantity
\begin{equation}
  \alpha_\mathrm{s} = \langle X \rangle + \frac{\Delta W}{\sqrt{8k} \Delta t} ,
\label{alphastochastic}
\end{equation}
where $\Delta W$ is a zero-mean, Gaussian random variable with
variance $\Delta t$. This alternate representation
as a stochastic variable will be useful later.
Since it will be clear from context, we will use $\alpha$ interchangeably
with $\alpha_\mathrm{s}$ in referring to the measurement results,
although technically we should distinguish between the index $\alpha$
and the stochastic variable $\alpha_\mathrm{s}$.

A \textit{continuous} measurement results if we make a sequence of these
measurements and take the limit as $\Delta t \longrightarrow 0$ (or
equivalently, as $\Delta t \longrightarrow dt$). As this limit is
taken, more and more measurements are made in any finite time
interval, but each is increasingly weak. By choosing the variance
of the measurement result to scale as $\Delta t$, we have ensured
that we obtain a sensible continuum limit. 
A stochastic equation of motion results due to the random
nature of the measurements (a \textit{stochastic} variable is
one that fluctuates randomly over time).
We can derive this
equation of motion for the system under this
continuous measurement by calculating the change induced in the
quantum state by the single weak measurement in the time step $\Delta t$,
to first order in $\Delta t$. 
We will thus compute the evolution when a measurement, 
represented by the operator $A(\alpha)$, is performed
in each time step.
This procedure gives
\begin{equation}
  \eqnarr{1.3}{
|\psi(t+\Delta t)\rangle &{}\propto{}&\ds  A(\alpha) |\psi(t)\rangle  \\
                         &{}\propto{}&\ds e^{-2k\Delta t(\alpha - X)^2} |\psi(t)\rangle  \\
                          &{}\propto{}&\ds e^{-2k\Delta t X^2 + X[ 4 k \langle X \rangle \Delta t + \sqrt{2k}\Delta W]} |\psi(t)\rangle .
  }
\end{equation}
We now expand the exponential to first order in $\Delta t$, which
gives
\begin{widetext}
\begin{equation}
  |\psi(t+\Delta t)\rangle \propto \{ 1 - 2k\Delta t X^2 + X[ 4 k \langle X \rangle \Delta t +
\sqrt{2k}\Delta W + kX(\Delta W)^2] \} |\psi(t)\rangle .
\end{equation}
\end{widetext}
Note that we have included the \textit{second-order} term in $\Delta W$ 
in the power series expansion
for the exponential.
We need
to include this term because it turns out that in the limit in which $\Delta t
\longrightarrow 0$, $(\Delta W)^2 \longrightarrow (dW)^2 = dt$. 
Because of this, the $(\Delta W)^2$ term contributes to the final differential 
equation. The reason for this will be explained in the next section, but for 
now we ask the reader to indulge us and accept that it is true. 

To take the limit as $\Delta t\rightarrow 0$, we set $\Delta t = dt$, 
$\Delta W = dW$ and $(\Delta W)^2 = dt$, and the result is
\begin{equation}
  |\psi(t+dt)\rangle \propto  \{ 1 -  [k X^2 - 4k X\langle X \rangle]\, dt + \sqrt{2k} X \,dW \}  |\psi(t)\rangle .
   \label{unnorm}
\end{equation}
This equation does not preserve the norm $\langle\psi|\psi\rangle$ 
of the wave function,
because before we derived it we threw away the normalization. We
can easily obtain an equation that \textit{does} preserve the norm simply
by normalizing $|\psi(t+dt)\rangle$ and expanding the result to
first order in $dt$ (again, keeping terms to order $dW^2$). 
Writing $|\psi(t+dt)\rangle = |\psi(t)\rangle
+ d|\psi\rangle$, the resulting stochastic differential equation
is given by
\begin{equation}
 d|\psi\rangle =  \{ - k (X - \langle X\rangle)^2 dt + \sqrt{2k} (X - \langle X\rangle)\, dW \}  |\psi(t)\rangle .
  \label{firstsse}
\end{equation}
This is the equation we have been seeking---it describes the
evolution of the state of a system in a time interval $dt$ given that the
observer obtains the measurement result
\begin{equation}
   dy = \langle X \rangle\, dt + \frac{dW}{\sqrt{8k}}
\label{positionmeasresult}
\end{equation} 
in that time interval. 
The measurement result gives the expected value $\,\expct{X}$ plus a random
component due to the width of $P(\alpha)$, and we write this as a differential
since it corresponds to the information gained in the time interval $dt$.
As the observer integrates $dy(t)$ the quantum state progressively collapses, and 
this integration is equivalent to solving (\ref{firstsse}) for the quantum-state evolution.

The \textit{stochastic Schr\"odinger equation}
(SSE) in Eq.~(\ref{firstsse}) is usually described as giving the
evolution {\em conditioned} upon the stream of measurement
results.
The state $\ket{\psi}$ evolves randomly, and 
$\ket{\psi(t)}$ is called the \textit{quantum trajectory} \cite{Carm93}.  
The set of measurement results $dy(t)$ is called the \textit{measurement record}.
We can also write this SSE in terms of the density operator
$\rho$ instead of $|\psi\rangle$. Remembering that we must keep
all terms proportional to $dW^2$, and defining $\rho(t+dt) \equiv
\rho(t) + d\rho$, we have
\begin{equation}
  \eqnarr{1.5}{
   d\rho \arreq (d|\psi\rangle ) \langle\psi | + |\psi\rangle (d\langle\psi | ) + (d|\psi\rangle )(d\langle\psi | )     \\
             \arreq - k[X[X,\rho]]\, dt \\
             &&\ds {}+ \sqrt{2k}(X\rho + \rho X - 2\langle X\rangle \rho) dW .
  }
 \label{SME}
\end{equation}
This is referred to as a \textit{stochastic master equation} (SME),
which also defines a quantum trajectory $\rho(t)$. This SME was first 
derived by Belavkin~\cite{BelavkinLQG}.  Note that in general, 
the SME also includes a term describing Hamiltonian evolution as in 
Eq.~(\ref{schrodvonneuman}).

The density operator at time $t$ gives the observer's state of knowledge
of the system, given that she has obtained the measurement record
$y(t)$ up until time $t$.  Since the observer has access to $dy$ but
not to $dW$, to calculate $\rho(t)$ she must calculate $dW$ at each
time step from the measurement record in that time step along with the
expectation value of $X$ at the previous time:
\begin{equation}
    dW =  \sqrt{8k} \, (dy -  \langle X \rangle \,dt).
\label{calculatedw}
\end{equation}
By substituting this expression in the SME [Eq.~(\ref{SME})], we can write the evolution of the 
system directly in terms of the measurement record, which is the natural 
thing to do from the point of the view of the observer. This is 
\begin{equation}
  \eqnarr{1.5}{
   d\rho   \arreq - k[X[X,\rho]]\, dt \\
             &&\ds {}+ 4k(X\rho + \rho X - 2\langle X\rangle \rho) (dy -  \langle X \rangle\, dt ).
  }
 \label{SMEwithdy}
\end{equation}

In Section~\ref{section:solution} we will explain how to solve the SME
analytically in a special case, but it is often necessary to solve it
numerically.  The simplest method of doing this is to take small time
steps $\Delta t$, and use a random number generator to select a new
$\Delta W$ in each time step.  One then uses $\Delta t$ and $\Delta W$
in each time step to calculate $\Delta\rho$ and adds this to the current
state $\rho$.  In this way we generate a specific trajectory for the
system.  Each possible sequence of $dW$'s generates a different
trajectory, and the probability that a given trajectory occurs is the
probability that the random number generator gives the corresponding
sequence of $dW$'s.  A given sequence of $dW$'s is often referred to
as a ``realization'' of the noise, and we will refer to the process of
generating a sequence of $dW$'s as ``picking a noise realization''.
Further details regarding the numerical methods for solving stochastic
equations are given in~\cite{Kloeden92}.

If the observer makes the continuous measurement, but throws away
the information regarding the measurement results, 
the observer must average over the different possible results.
Since $\rho$ and $dW$ are statistically independent, $\dexpct{\rho\, dW} =
0$, where the double brackets denote this average (as we show in 
Section~\ref{section:ensembleaveragees}).
The result is thus given by setting to zero all terms
proportional to $\rho\,dW$ 
in Eq.~(\ref{SME}),
\begin{equation}
 \frac{d\rho}{dt} = - k[X[X,\rho]] ,
 \label{ME}
\end{equation}
where the density operator here represents the state averaged over
all possible measurement results.
We note that the method we have used above to derive the stochastic Schr\"odinger 
equation is an extension of a method initially developed by Caves and Milburn to 
derive the (non-stochastic) master equation (\ref{ME})~\cite{CMnotes}.

\section{An Introduction to Stochastic Calculus}
\label{section:Ito}

Now that we have encountered a noise process in the quantum evolution,
we will explore in more detail the formalism for handling this.
It turns out that adding a white-noise stochastic process changes the
basic structure of the calculus for treating the evolution equations.
There is more than one formulation 
to treat stochastic 
processes, but the one referred to as \textit{It\^o calculus} is used in almost 
all treatments of noisy quantum systems, and so this is the one we describe 
here. The main alternative formalism may be found in
Refs.~\cite{gardiner,Kloeden92}.

\subsection{Usage}

First, let's review the usual calculus in a slightly different way.
A differential equation
\begin{equation}
  \frac{dy}{dt} = \alpha
\end{equation}
can be instead written in terms of differentials as
\begin{equation}
   dy= \alpha\,dt.
\end{equation}
The basic rule in the familiar \textit{deterministic} calculus is that $(dt)^2=0$.
To see what we mean by this, we can try calculating the differential $dz$ for 
the variable $z=e^y$ in terms of the differential for $dy$  as follows:
\begin{equation}
  dz = e^{y+dy}-e^y = z\left(e^{\alpha\,dt}-1\right).
\end{equation}
Expanding the exponential and applying the rule $(dt)^2=0$,
we find
\begin{equation}
  dz = z\alpha\,dt.
\end{equation}
This is, of course, the same result as that obtained by using the chain rule 
to calculate $dz/dy$ and multiplying through by $dy$.
The point here is that calculus breaks up functions and considers
their values within short intervals $\Delta t$.  In the infinitesimal
limit, the quadratic and higher order terms in $\Delta t$ end up being
too small to contribute.

In It\^o calculus, we have an additional differential element $dW$,
representing white noise.  The basic rule of It\^o calculus is
that $dW^2=dt$, while $dt^2=dt\,dW = 0$.  We will justify this
later, but to use this calculus, we simply note that we ``count''
the increment $dW$ as if it were equivalent to $\sqrt{dt}$ in
deciding what orders to keep in series expansions
of functions of $dt$ and $dW$.
As an example, consider the stochastic differential equation
\begin{equation}
  dy = \alpha\,dt+\beta\,dW.
  \label{exampley}
\end{equation}
We obtain the corresponding differential equation for $z=e^y$
by expanding to \textit{second} order in $dy$:
\begin{equation}
  dz = e^y\left(e^{dy}-1\right) = z\left(dy+\frac{(dy)^2}{2}\right).
\end{equation}
Only the $dW$ component contributes to the quadratic term; the
result is
\begin{equation}
  dz = z\left(\alpha+ \frac{\beta^2}{2}\right)\,dt + z\beta\,dW.
\end{equation}
The extra $\beta^2$ term is crucial in understanding many phenomena
that arise in continuous-measurement processes.

\subsection{Justification}

\subsubsection{Wiener Process}\label{section:wienerprocess}
To see why all this works, let's first define the \textit{Wiener process}
$W(t)$ as an ``ideal'' random walk with arbitrarily small, independent 
steps taken arbitrarily often.  (The Wiener process is thus
scale-free and in fact fractal.)  Being a symmetric random walk, $W(t)$ is 
a normally distributed random variable with zero mean, 
and we choose
the variance of $W(t)$ to be $t$ (i.e., the width of the 
distribution is $\sqrt{t}$, as is characteristic of a diffusive process).
We can thus write the probability density for $W(t)$ as
\begin{equation}
  P(W,t) = \frac{1}{\sqrt{2\pi t}} e^{-W^2/2t}.
  \label{gaussprobW}
\end{equation}
In view of the central-limit theorem, \textit{any} simple random
walk gives rise to a Wiener process in the continuous limit, independent
of the one-step probability distribution (so long as the one-step
variance is finite).

Intuitively, $W(t)$ is a continuous but everywhere nondifferentiable function.
Naturally, the first thing we will want to do is to develop the analogue
of the derivative for the Wiener process.
We can start by defining the Wiener increment
\begin{equation}
  \Delta W(t) := W(t+\Delta t)-W(t)
\end{equation}
corresponding to a time increment $\Delta t$.
Again, $\Delta W$ is a normally distributed random variable with 
zero mean and variance $\Delta t$.
Note again that this implies that the root-mean-square amplitude of
$\Delta W$ scales as $\sqrt{\Delta t}$.  We can understand this
intuitively since the \textit{variances} add for successive steps
in a random walk.  Mathematically, we can write the variance as
\begin{equation}
  \dexpct{(\Delta W)^2} = \Delta t,
\end{equation}
where the double angle brackets $\dexpct{~}$ denote an ensemble
average over all possible realizations of the Wiener process.
This relation suggests the above notion that second-order terms in
$\Delta W$ contribute at the same level as first-order terms in $\Delta t$.
In the infinitesimal limit of $\Delta t\longrightarrow 0$, we 
write $\Delta t\longrightarrow dt$
and $\Delta W\longrightarrow dW$.

\subsubsection{It\^o Rule}

We now want to show that the Wiener differential $dW$ satisfies 
the It\^o rule $dW^2=dt$.  Note that we want this to hold 
\textit{without} the ensemble average, which is surprising since
$dW$ is a stochastic quantity, while $dt$ obviously is not.
To do this, consider the probability density function for $(\Delta W)^2$,
which we can obtain by a simple transformation of the Gaussian
probability density for $\Delta W$ [which is Eq.~(\ref{gaussprobW})
with $t\longrightarrow\Delta t$ and $W\longrightarrow\Delta W$]:
\begin{equation}
  P\left[(\Delta W)^2\right] = 
   \frac{e^{-(\Delta W)^2/2\Delta t}}{\sqrt{2\pi\,\Delta t\,(\Delta W)^2}}.
\end{equation}
In particular, the mean and variance of this distribution for $(\Delta W)^2$ are
\begin{equation}
  \dexpct{(\Delta W)^2} = \Delta t
\end{equation}
and 
\begin{equation}
  \textrm{Var}\left[(\Delta W)^2\right] = 2(\Delta t)^2,
\end{equation}
respectively.
To examine the continuum limit, we will sum the Wiener increments over
$N$ intervals of duration $\Delta t_N = t/N$ between $0$ and $t$. 
The corresponding Wiener increments are
\begin{equation}
  \Delta W_n := W[(n+1)\Delta t_N] - W(n\Delta t_N).
\end{equation}
Now consider the sum of the squared increments
\begin{equation}
  \sum_{n=0}^{N-1}(\Delta W_n)^2,
  \label{W2sum}
\end{equation}
which corresponds to a random walk of $N$ steps, where a single
step has average value $t/N$ and variance $2t^2/N^2$.
According to the central limit theorem, for large $N$ the
sum (\ref{W2sum}) is a Gaussian random variable with mean $t$ and
variance $2t^2/N$.
In the limit $N\longrightarrow \infty$, the variance of the
sum vanishes, and the sum becomes $t$ with certainty.
Symbolically, we can write
\begin{equation}
  \int_0^t [dW(t')]^2 :=\lim_{N\rightarrow\infty}\sum_{n=0}^{N-1}
     (\Delta W_n)^2 = t = \int_0^t dt'.
\end{equation}
For this to hold over any interval $(0,t)$, we must make the 
formal identification $dt=dW^2$.
This means that even though $dW$ is a random variable, $dW^2$ is not,
since it has no variance when integrated over any finite interval.

\subsubsection{Ensemble Averages}
\label{section:ensembleaveragees}

Finally, we need to justify a relation useful for averaging over
noise realizations, namely that
\begin{equation}
  \dexpct{y\,dW}=0
  \label{vanishexpect}
\end{equation} 
for a solution $y(t)$ of Eq.~(\ref{exampley}). This makes it particularly 
easy to compute averages of functions of $y(t)$ over all possible 
realizations of a Wiener process, 
since we can simply set $dW=0$, even when it is multiplied 
by $y$.  We can see this as follows. Clearly, $\dexpct{dW}=0$. Also, 
Eq.~(\ref{exampley}) is the continuum limit of the discrete relation
\begin{equation}
  y(t+\Delta t) = y(t) +  \alpha\Delta t + \beta\Delta W(t).
\end{equation}
Thus, $y(t)$ depends on $\Delta W(t-\Delta t)$, but is
independent of $W(t)$, which gives the desired result,
Eq.~(\ref{vanishexpect}).  More detailed discussions of Wiener 
processes and It\^o calculus may be found in~\cite{WienerIntroPaper,gardiner}

\section{Solution of a Continuous Measurement}
\label{section:solution}

The stochastic equation (\ref{SME}) that describes the
dynamics of a system subjected to a continuous measurement is
nonlinear in $\rho$, which makes it difficult to solve. However, it turns
out that this equation can be recast in an effectively
equivalent but linear form. We now derive this linear form, and then show 
how to use it to obtain a complete solution to the SME. To do this, we first 
return to the unnormalized stochastic Schr\"{o}dinger 
equation (\ref{unnorm}). Writing this in terms of the measurement 
record $dy$ from Eq.~(\ref{positionmeasresult}), we have 
\begin{equation}
  |\tilde{\psi}(t+dt)\rangle =  \{ 1 -  k X^2 dt + 4kX\, dy  \}  |\tilde{\psi}(t)\rangle ,
  \label{unnorm2}
\end{equation}
where the tilde denotes that the state is not normalized (hence the 
equality here). 
Note that the nonlinearity 
in this equation is entirely due to the fact that 
$dy$ depends upon $\langle X \rangle$ 
(and $\langle X \rangle$ depends upon $\rho$). So what would 
happen if we simply replaced $dy$ in this equation with $dW/\sqrt{8k}$? 
This would mean that we would be choosing the measurement record
incorrectly in each time step $dt$.  
But the ranges of both $dy$ and $dW$ are the full real line, 
so replacing $dy$ by $dW/\sqrt{8k}$ still corresponds to 
a \textit{possible} realization of $dy$.  However, 
we would then be using the wrong \textit{probability density}
for $dy$ because $dy$ and $dW/\sqrt{8k}$ have different means.
Thus, if we were to use $dW/\sqrt{8k}$ in place of $dy$
we would obtain all the correct trajectories, but with the wrong
probabilities.

Now recall from Section~\ref{section:weak} that when we apply a
measurement operator to a quantum state, we must explicitly
renormalize it.  If we don't renormalize, the new norm contains
information about the prior state: it represents the 
\textit{prior} probability that the particular measurement outcome
actually occured.
Because the operations that result in each
succeeding time interval $dt$ are independent, and probabilities
for independent events multiply, this statement remains true after any number of
time steps.  That is, after $n$ time steps, the norm of the state
records the probability that the sequence of measurements led to that
state.  To put it yet another way, it records the probability that
that particular trajectory occurred.  This is extremely useful,
because it means that we do not have to choose the trajectories with
the correct probabilities---we can recover these at the end merely
by examining the final norm! 

To derive the linear form of the SSE we use the observations above.
We start with the {\em normalized} form given by Eq.~(\ref{firstsse}), and 
write it in terms of $dy$, which gives
\begin{equation}
 \eqnarr{1.5}{
  |\psi(t+dt)\rangle \arreq  \{ 1 - k (X - \langle X\rangle)^2 dt \\
    &&\ds {}+ 4k (X - \langle X\rangle) (dy - \langle X \rangle \,dt)  \}  |\psi(t)\rangle .
  }
\end{equation}
We then replace $dy$ by $dW/\sqrt{8k}$ (that is, we remove the mean from
$dy$ at each time step).  In addition, we multiply the state by the square root of the 
actual probability for getting that state (the probability for $dy$)
and divide by the square root of the probability for $dW$.  To first order in $dt$, the factor we multiply by is therefore 
\begin{equation}
  \sqrt{\frac{P(dW)}{P(dy)}} = 1 + \sqrt{2k}\langle X\rangle dW - k \langle X\rangle^2 dt .
\end{equation}
The resulting stochastic equation is linear, being 
\begin{equation}
  |\tilde{\psi}(t+dt)\rangle =  \{ 1 -  k X^2 dt + \sqrt{2k}X\, dW  \}  |\tilde{\psi}(t)\rangle . 
  \label{linSSE}
\end{equation}
The linear stochastic master equation equivalent to this linear SSE is 
\begin{equation}
 d\tilde{\rho} = - k[X[X,\tilde{\rho}]] dt + \sqrt{2k}(X\tilde{\rho} + \tilde{\rho} X) dW.
 \label{lin}
\end{equation}
Because of the way we have constructed this equation,  the
actual probability at time $t$
for getting a particular trajectory is the product of (1)
the norm of the state at time $t$ and (2)
the probability that the trajectory is generated by the linear
equation (the latter factor being the probability for picking
the noise realization that generates the trajectory.)  
This may sound complicated, but it is actually quite simple in practice, as we will
now show.  Further information regarding linear SSE's may be found in
the accessible and detailed discussion given by Wiseman
in~\cite{WisemanLinQ}.

We now solve the linear SME to obtain a complete 
solution to a quantum measurement in the special case in which the Hamiltonian 
commutes with the measured observable $X$. A
technique that allows a solution to be obtained in some more
general cases may be found in Ref.~\cite{JK}. 
To solve Eq.~(\ref{lin}), we include a Hamiltonian of the form $H =
f(X)$, and write the equation as an exponential to first order in
$dt$. The result is
\begin{equation}
  \eqnarr{1.5}{
   \tilde{\rho}(t+dt) \arreq e^{[-iH/\hbar  - 2k X^2]dt + \sqrt{2k} X
   dW} \tilde{\rho}(t)  \\
                & &\ds {}\times e^{[iH/\hbar  - 2k X^2]dt + \sqrt{2k} X dW},
  }
\end{equation}
which follows by expanding the exponentials (again to first order in 
$dt$ and second order in $dW$) to see that this expression is equivalent
Eq.~(\ref{lin}).
What we have written is the generalization of the usual
unitary time-evolution operator under standard
Schr\"odinger-equation evolution.
The evolution for a finite time $t$ is easily obtained now by
repeatedly multiplying on both sides by these exponentials. We can
then combine all the exponentials on each side in a single
exponential, since all the operators commute. The result is
\begin{equation}
  \eqnarr{1.5}{
   \tilde{\rho}(t;W) \arreq e^{[-iH/\hbar - 2k X^2]t + \sqrt{2k} X W}
   \tilde{\rho}(0)  \\
                & &\ds {}\times e^{[i H/\hbar - 2k X^2]t + \sqrt{2k} X W} ,
   }
\end{equation}
where the final states $\tilde\rho(t;W)$ are parameterized by $W$, with
\begin{equation}
  W = \int_{0}^{t} dW(t') .
\end{equation}
The probability density for $W\!$, being the sum of the Gaussian
random variables $dW\!$, is Gaussian. In particular, 
as in Eq.~(\ref{gaussprobW}),  
at time $t$ the probability density is
\begin{equation}
  \tilde{P}(W,t) = \frac{1}{\sqrt{2\pi t}} e^{-W^2/(2t)} .
\end{equation}
That is, at time $t$, $W$ is a Gaussian random variable with mean
zero and variance $t$.

As we discussed above, however, the probability for obtaining
$\rho(t)$ is not the probability with which it is generated by
picking a noise realization. To calculate the ``true'' probability for
$\rho(t)$ we must multiply the density $P(W,t)$ by the norm of
$\tilde\rho(t)$. Thus, the actual probability for getting a final
state $\rho(t)$ (that is, a specific value of $W$ at time $t$) is
\begin{equation}
  P(W,t) = \frac{1}{\sqrt{2\pi t}} e^{-W^2/(2t)}
           \mbox{Tr}\left[e^{[- 4k X^2]t + \sqrt{8k} X W}\rho(0)\right] .
\end{equation}

\begin{figure}[tb]
  \begin{center}
     \includegraphics[width=1.0\hsize]{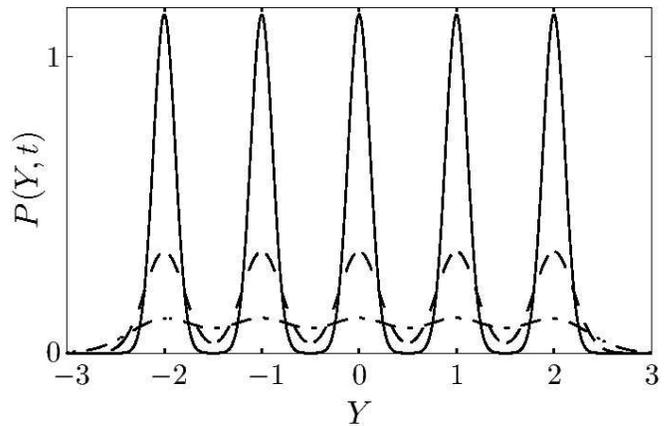}
  \end{center}
  \vspace{-5mm}
  \caption
         {Here we show the probability density for the result of a measurement 
          of the $z$-component of angular momentum for $j=2$, and with 
          measurement strength $k$. This density 
          is shown for three different measurement times: dot-dashed line: $t = 1/k$; 
          dashed line: $t=3/k$; solid line: $t = 10/k$. 
	\label{fig:msolution}}
\end{figure}

At this point, $X$ can just as well be \textit{any} Hermitian operator. Let us now assume
that $X = J_z$ for some quantum number $j$ of the angular momentum. In this
case $X$ has $2j + 1$ eigenvectors $|m\rangle$, with
eigenvalues $m = -j,-j+1, \ldots, j$. If 
we have no information about the system
at the start of the measurement,
so that the
initial state is $\rho(0) = I/(2j+1)$, then the solution is quite
simple. In particular, $\rho(t)$ is diagonal in the $J_z$
eigenbasis, and
\begin{equation}
  \langle m |\rho(t)| m \rangle = \frac{e^{-4kt(m-Y)^2}}{\cal N}
\end{equation}
where ${\cal N}$ is the normalization and 
$Y:=W/(\sqrt{8k}\,t)$. The true probability density for $Y$ is
\begin{equation}
  P(Y,t) = \frac{1}{2j+1}\sum_{n=-j}^j \sqrt{\frac{4kt}{\pi}}\,e^{-4kt(Y-n)^2} .
\end{equation}
We therefore see that after a sufficiently long time, the density
for $Y$ is sharply peaked about the $2j+1$ eigenvalues of $J_z$. 
This density is plotted in Fig.~\ref{fig:msolution} for three values of $t$.
At long times, $Y$ becomes very close to one of these
eigenvalues. Further, we see from the solution for $\rho(t)$ that
when $Y$ is close to an eigenvalue $m$, then the state of the
system is sharply peaked about the eigenstate $|m\rangle$. Thus, we
see that after a sufficiently long time, the system is projected
into one of the eigenstates of $J_z$.

The random variable $Y$ has a physical meaning.  Since we replaced the
measurement record $dy$ by $dW/\sqrt{8k}$ to obtain the linear
equation, when we transform from the raw probability density
$\tilde{P}$ to the true density $P$ this transforms the driving noise
process $dW$ back into $\sqrt{8k}\, dy = \sqrt{8k} \langle X(t)\rangle
dt + dW$, being a scaled version of the measurement record.  Thus,
$Y(t)$, as we have defined it, is actually the output record up until
time $t$, divided by $t$.  That is,
\begin{equation}
  Y = \frac{1}{t}\int_0^t \langle J_z(t) \rangle dt +
      \frac{1}{\sqrt{8k}\,t}\int_0^t dW .
\end{equation}
Thus, $Y$ is the measurement result. When making the measurement
the observer integrates up the measurement record, and then
divides the result by the final time. The result is $Y$, and the
closer $Y$ is to one of the eigenvalues, and the longer the time
of the measurement, the more certain the observer is that the
system has been collapsed onto the eigenstate with that
eigenvalue.
Note that as the measurement progresses, the second, explicitly stochastic
term converges to zero, while the expectation value in the first term
evolves to the measured eigenvalue.

\section{Multiple Observers and Inefficient Detection}
\label{section:multipleobservers}

It is not difficult to extend the above analysis to describe what
happens when more than one observer is monitoring the system.
Consider two observers Alice and Bob, who measure the same system.
Alice monitors $X$ with strength $k$, and Bob monitors $Y$ with
strength $\kappa$.  From Alice's point of view, since she has no
access to Bob's measurement results, she must average over them.  Thus,
as far as Alice is concerned, Bob's measurement simply induces the
dynamics $d\rho_1 = -\kappa [Y,[Y,\rho_1]]$ where $\rho_1$ is her
state of knowledge.  The full dynamics of her state of knowledge,
including her measurement, evolves according to
\begin{equation}
  \eqnarr{1.5}{
   d\rho_1 \arreq  - k[X[X,\rho_1]]dt  - \kappa[Y[Y,\rho_1]]dt   \\
            &&\ds {}+ \sqrt{2k}(X\rho_1 + \rho_1 X - 2\langle X\rangle_1 \rho_1) dW_1 ,
  }
\end{equation}
where $\expct{X}_1:=\mathrm{Tr}[X\rho_1]$, 
and her measurement record is $dr_1 = \expct{X}_1 dt + dW_1/\sqrt{8k}$. 
Similarly, the equation of motion for Bob's state of knowledge is 
\begin{equation}
  \eqnarr{1.5}{
   d\rho_2 \arreq  - \kappa[Y[Y,\rho_2]]dt - k [X[X,\rho_2]]dt  \\
            &&\ds + \sqrt{2\kappa}(Y\rho_2 + \rho_2 Y - 2\langle Y\rangle_2 \rho_2) dW_2 ,
  }
\end{equation}
and his measurement record is $dr_2 = \expct{Y}_2 dt + dW_2/\sqrt{8\kappa}$.

We can also consider the state of knowledge of a single observer,
Charlie, who has access to both measurement records $dr_1$ and
$dr_2$.  The equation for Charlie's state of knowledge, $\rho$, is
obtained simply by applying both measurements simultaneously, giving
\begin{equation}
  \eqnarr{1.5}{
   d\rho \arreq  - k[X[X,\rho]]dt + \sqrt{2k}\left(X\rho + \rho X - 2\langle X\rangle \rho\right) dV_1 \\
            & &\ds {} -  \kappa[Y[Y,\rho]]dt + \sqrt{2\kappa}\left(Y\rho + \rho Y - 2\langle Y\rangle \rho\right) dV_2 ,
  }
\end{equation}
where $\expct{X}:=\mathrm{Tr}[X\rho]$.
Note that $dV_1$ and $dV_2$ are independent noise sources.  In terms
of Charlie's state of knowledge the two measurement records are
 \begin{equation}
  \eqnarr{2.0}{
   dr_1 \arreq \expct{X} dt + \frac{dV_1}{\sqrt{8k}}  , \\
   dr_2 \arreq  \expct{Y} dt + \frac{dV_2}{\sqrt{8\kappa}}  .
  }
\end{equation}
In general Charlie's state of knowledge $\rho(t) \not= \rho_1(t) \not=
\rho_2(t)$, but Charlie's measurement records are the same as
Alice's and Bob's.  Equating Charlie's expressions for the measurement
records with Alice's and Bob's, we obtain the relationship between
Charlie's noise sources and those of Alice and Bob:
 \begin{equation}
  \eqnarr{1.5}{
   dV_1 \arreq \sqrt{8k} \left(\,\expct{X}_1 - \expct{X}\right)dt + dW_1 , \\
   dV_2 \arreq \sqrt{8\kappa} \left(\,\expct{Y}_2 - \expct{Y}\right)dt + dW_2 .
  }
\end{equation}
We note that in quantum optics, each measurement is often referred to
as a separate ``output channel'' for information, and so multiple
simultaneous measurements are referred to as multiple output channels.
Multiple observers were first treated explicitly by Barchielli, who
gives a rigorous and mathematically sophisticated
treatment in Ref.~\cite{Barchielli93}.  A similarly detailed and considerably
more accessible treatment is given in Ref.~\cite{Dziarmaga04}.

We turn now to {\em inefficient} measurements, which can be treated in
the same way as multiple observers.  An inefficient measurement is one
in which the observer is not able to pick up all the measurement
signal.  The need to consider inefficient measurements arose
originally in quantum optics, where photon counters will only detect
some fraction of the photons incident upon them.  This fraction,
usually denoted by $\eta$, is referred to as the \textit{efficiency of the
detector}~\cite{WM93}.  A continuous measurement in which the detector
is inefficient can be described by treating the single measurement as
two measurements, where the strengths of each of them sum to the
strength of the single measurement.  Thus we rewrite the equation for
a measurement of $X$ at strength $k$ as
\begin{equation}
  \eqnarr{1.5}{
   d\rho \arreq  - k_1[X[X,\rho]]\,dt + \sqrt{2k_1}(X\rho + \rho X - 2\langle X\rangle \rho)\, dV_1 \\
            &&\ds  - k_2[X[X,\rho]]\,dt + \sqrt{2k_2}(X\rho + \rho X - 2\langle X\rangle \rho)\, dV_2 ,
  }
\end{equation}
where $k_1 + k_2 = k$.  We now give the observer access to only the
measurement with strength $k_1$.  From our discussion above, the
equation for the observer's state of knowledge, $\rho_1$, is
\begin{equation}
  \eqnarr{1.5}{
   d\rho_1 \arreq  - (k_1 + k_2) [X[X,\rho_1]]\,dt  \\ 
                &&\ds {} + \sqrt{2k_1}(X\rho_1 + \rho_1 X - 2\langle X\rangle_1 \rho_1)\, dW_1 \\
            \arreq - k[X[X,\rho_1]]\,dt +  \\ 
            &&\ds \sqrt{2\eta k}(X\rho_1 + \rho_1 X - 2\langle X\rangle_1 \rho_1)\, dW_1 ,
  }
\end{equation}
where, as before, the measurement record is
\begin{equation}
  dr_1 = \expct{X}_1 dt + \frac{dW_1}{\sqrt{8k_1}} 
       = \expct{X}_1 dt + \frac{dW_1}{\sqrt{8\eta k}},
  \label{ineffposmeas}
\end{equation}
and
\begin{equation}
  \eta = \frac{k_1}{k_1 + k_2} = \frac{k_1}{k}
\end{equation}
is the efficiency of the detector.

\section{General Form of the Stochastic Master Equation}
\label{section:genform}

Before looking at a physical example of a continuous measurement process,
it is interesting to ask, what is the most general form of the measurement 
master equation when the measurements involve Gaussian noise?  
In this section we present a simplified version of an
argument by Adler~\cite{Adler00} that allows one to derive a form 
that is close to the fully general one and sufficient for most purposes. We 
also describe briefly the extension that gives the fully general form, the details 
of which have been worked out by Wiseman and Diosi~\cite{Wiseman01}. 

Under unitary (unconditioned) evolution, the Schr\"odinger equation
tells us that in a short time interval $dt$, the state vector
undergoes the transformation 
\begin{equation}
  \ket\psi \longrightarrow \ket\psi + d\ket\psi =
    \left(1-i\frac{H}{\hbar}\,dt\right)\ket\psi,
\end{equation}
where $H$ is the Hamiltonian.
The same transformation applied to the density operator gives the
Schr\"odinger--von Neumann equation of Eq.~(\ref{schrodvonneuman}):
\begin{equation}
    \rho + d\rho = 
    \left(1-i\frac{H}{\hbar}\,dt\right)\rho \left(1+i\frac{H}{\hbar}\,dt\right)
    = \rho-\frac{i}{\hbar}[H,\rho]\,dt.
\end{equation}
To be physical, any transformation of the density operator must be
\textit{completely positive}. That is, the transformation must preserve the
fact that the density operator has only nonnegative eigenvalues.
This property guarantees that the density operator can generate only sensible 
(nonnegative)
probabilities. (To be more precise, \textit{complete} positivity means that 
the transformation for a system's density operator 
must preserve the positivity of the density operator---the fact that
the density operator has no negative eigenvalues---of any larger system 
containing
the system~\cite{mikeandike}.)
It turns out that the most general form of a completely positive
transformation is
\begin{equation}
  \rho\longrightarrow \sum_n A_n\rho A_n^\dagger,
  \label{rhotransform}
\end{equation}
where the $A_n$ are arbitrary operators.
The Hamiltonian evolution above corresponds to a single infinitesimal
transformation
operator $A=1-iH\,dt/\hbar$.

Now let's examine the transformation for a more general, \textit{stochastic}
operator of the form
\begin{equation}
  A = 1-i\frac{H}{\hbar}\,dt + b\,dt + c\, dW,
\end{equation}
where $b$ and $c$ are operators.
We will use this operator to ``derive'' a Markovian master equation, then 
indicate how it can be made more general.
We may assume here that $b$ is Hermitian, since we can absorb any antihermitian
part into the Hamiltonian.
Putting this into the transformation (\ref{rhotransform}), we find
\begin{equation}
  d\rho = -\,\frac{i}{\hbar}[H,\rho]\,dt + [b, \rho]_+dt + c\rho c^\dagger\,dt
   + \left(c\rho+\rho c^\dagger\right)\, dW,
  \label{masteqbc}
\end{equation}
where $[A,B]_+:=AB+BA$ is the anticommutator.
We can then take an average over all possible Wiener processes, which
again we denote by the double angle brackets 
$\dexpct{~}$.  From Eq.~(\ref{vanishexpect}), $\dexpct{\rho\,dW}=0$ in It\^o calculus, so
\begin{equation}
  d\dexpct{\rho} = -\,\frac{i}{\hbar}\left[H,\dexpct{\rho}\right]\,dt
     + \left[b,\dexpct{\rho}\right]_+dt + c\dexpct{\rho}\!c^\dagger\,dt.
\end{equation}
Since the operator $\dexpct{\rho}$ is an average over valid density
operators, it is also a valid density operator and must therefore
satisfy $\mathrm{Tr}[\dexpct{\rho}]=1$.  Hence we must have
$d\mathrm{Tr}[\dexpct{\rho}]= \mathrm{Tr}[d\!\dexpct{\rho}]=0$.  Using
the cyclic property of the trace, this gives
\begin{equation}
  \mathrm{Tr}\left[\dexpct{\rho}\left(2b+c^\dagger c\right)\right]=0.
\end{equation}
This holds for an arbitrary density operator only if
\begin{equation}
  b=-\,\frac{c^\dagger c}{2}.
\end{equation}
Thus we obtain the \textit{Lindblad form}~\cite{Lindblad76} of the
master equation (averaged over all possible noise realizations):
\begin{equation}
  d\dexpct{\rho} = -\,\frac{i}{\hbar}\left[H,\dexpct{\rho}\right]\,dt
     + \mathcal{D}[c]\!\dexpct{\rho}\,dt.
\end{equation}
Here, we have defined the Lindblad superoperator
\begin{equation}
  \mathcal{D}[c]\rho := c\rho c^\dagger - \frac{1}{2}\left(c^\dagger c\rho + \rho c^\dagger c\right),
  \label{lindbladsuperop}
\end{equation}
where ``superoperator'' refers to the fact that $\mathcal{D}[c]$ operates 
on $\rho$ from both sides.
This is the most general (Markovian) form of the unconditioned 
master equation for a single dissipation process.

The full transformation from Eq.~(\ref{masteqbc}) then becomes
\begin{equation}
  d\rho = -\,\frac{i}{\hbar}[H,\rho]\,dt + \mathcal{D}[c]\rho \,dt
   + \left(c\rho+\rho c^\dagger\right)\, dW. 
\end{equation}
This is precisely the linear master equation, for which we already
considered the special case of $c=\sqrt{2k}X$ for the measurement
parts in Eq.~(\ref{lin}).
Again, this form of the master equation does not in general preserve the trace of the
density operator, since the condition $\mathrm{Tr}[d\rho]=0$ implies
\begin{equation}
  \mathrm{Tr}\left[\rho\left(c+c^\dagger\right)\,dW\right]=0.
  \label{traceccdag}
\end{equation}
We could interpret this relation as a constraint on $c$~\cite{Adler00}, 
but we will instead 
keep $c$ an arbitrary operator and explicitly renormalize
$\rho$ at each time step by adding a term proportional to the
left-hand side of (\ref{traceccdag}).
The result is the nonlinear form
\begin{equation}
  d\rho = -\,\frac{i}{\hbar}[H,\rho]\,dt + \mathcal{D}[c]\rho \,dt
   + \mathcal{H}[c]\rho\, dW,
   \label{gensme1}
\end{equation}
where the measurement superoperator is 
\begin{equation}
  \mathcal{H}[c]\rho := c\rho + \rho c^\dagger - \expct{c+c^\dagger}\rho.
\end{equation}
When $c$ is Hermitian, the measurement terms again give 
precisely the stochastic master equation~(\ref{SME}).

More generally, we may have any number of measurements, sometimes 
referred to as {\em output channels}, happening simultaneously. The result is 
\begin{equation}
  d\rho = -\,\frac{i}{\hbar}[H,\rho]\,dt + \sum_n \left(\mathcal{D}[c_n]\rho\,dt
   + \mathcal{H}[c_n]\rho\, dW_n\right).
\end{equation}
This is the same as Eq.~(\ref{gensme1}), but this time summed
(integrated) over multiple possible measurement operators $c_n$, each
with a separate Wiener noise process independent of all the others.  

In view of the arguments of Section~(\ref{section:multipleobservers}), when 
the measurements are inefficient, we have  
\begin{equation}
  d\rho = -\,\frac{i}{\hbar}[H,\rho]\,dt + \sum_n \left(\mathcal{D}[c_n]\rho\,dt,
   + \sqrt{\eta_n} \mathcal{H}[c_n]\rho\, dW\right), 
  \label{generalmasteqn}
\end{equation}
where $\eta_n$ is the efficiency of the $n$th detection channel.
The corresponding measurement record for the $n$th process
can be written
\begin{equation} 
  dr(t) = \frac{\expct{c_n+c_n^\dagger}}{2}dt + \frac{dW_n}{\sqrt{4 \eta_n}}.
\end{equation}
Again, for a single, position-measurement 
channel of the form $c=\sqrt{2k}X$, we recover  
Eqs.~(\ref{positionmeasresult}) and (\ref{ineffposmeas}) if we identify 
$dr_n/\sqrt{2k}$ as a rescaled measurement record.

The SME in Eq.~(\ref{generalmasteqn}) is sufficiently 
general for most purposes when one is concerned with measurements 
resulting in Wiener noise, but is not quite the most general form 
for an SME driven by such noise. The most general form 
is worked out in Ref.~\cite{Wiseman01}, and includes the fact 
that the noise sources may also be complex and mutually correlated.

\section{Interpretation of the Master Equation}
\label{section:interpretation}

Though we now have the general form of the master equation (\ref{generalmasteqn}),
the interpretation of each of the measurement terms is not entirely obvious.
In particular, the 
$\mathcal{H}[c]\rho$ terms (i.e., the noise terms) represent the 
information gain due to the measurement process, while the
$\mathcal{D}[c]\rho$ terms represent the disturbance to, or 
the \textit{backaction} on,
the state of the system
due to the measurement.  Of course, as we see from the 
dependence on the efficiency $\eta$, the backaction occurs
independently of whether the observer uses or discards the
measurement information (corresponding to $\eta=1$ or $0$, respectively).

To examine the roles of these terms further, we will now consider the
equations of motion for the moments (expectation values of powers
of $X$ and $P$) of the canonical variables.
In particular, we will specialize to the case of a single measurement channel,
\begin{equation}
  d\rho = -\,\frac{i}{\hbar}[H,\rho]\,dt + \mathcal{D}[c]\rho \,dt
   + \sqrt{\eta}\mathcal{H}[c]\rho\, dW.
\end{equation}
For an arbitrary operator $A$, we can use the master equation 
and $d\expct{A}=\mathrm{Tr}[A\,d\rho]$ to obtain following equation of motion
for the expectation value $\expct{A}$:
\begin{equation}
  \eqnarr{2.1}{
  d\expct{A} \arreq -\,\frac{i}{\hbar}\expct{[A,H]}\,dt \\ &&\ds
     +\expct{c^\dagger Ac-\frac{1}{2}\left(c^\dagger cA+Ac^\dagger c\right)}\,dt\\
   &&\ds {}+ \sqrt{\eta}\expct{c^\dagger A+Ac-\expct{A}\expct{c+c^\dagger}} \, dW.
  }
  \label{expctA}
\end{equation}
Now we will consider the effects of measurements on the relevant expectation
values in two example cases: a position measurement, corresponding to an observable, and
an antihermitian operator, corresponding to an energy damping process.
As we will see, the interpretation differs slightly in the two cases.
For concreteness and simplicity, we will assume the system is a
harmonic oscillator of the form
\begin{equation}
  H=\frac{P^2}{2m}+\frac{1}{2} m\omega_0^{\,2}X^2,
\end{equation}
and consider the lowest few moments of $X$ and $P$.  We will
also make the simplifying assumption that the initial
state is Gaussian, so that we only need to consider the simplest five 
moments: the means\, $\expct{X}$ and $\expct{P}$, the variances
$\Vx$ and $\Vp$, where $V_\alpha:= \expct{\alpha^2}-\expct{\alpha}^2$,
and the symmetrized covariance $\Cxp:=(1/2)\expct{[X,P]_+}-\expct{X}\expct{P}$.
These moments completely characterize arbitrary Gaussian states
(including mixed states).

\subsection{Position Measurement}

In the case of a position measurement of the form 
$c=\sqrt{2k}\,X$ as in Eq.~(\ref{lin}), Eq.~(\ref{expctA}) becomes
\begin{equation}
  \eqnarr{1.5}{
  d\expct{A} \arreq -\,\frac{i}{\hbar}\expct{[A,H]}\,dt -k\expct{[X,[X,A]]}\,dt\\
   &&\ds {}+ \sqrt{2\eta k}\left[\expct{[X,A]_+}-2\expct{X}\expct{A}\right]  \, dW.
  }
\end{equation}
Using this equation to compute the cumulant equations of motion, we find
\cite{DJ}
\begin{equation}
  \eqnarr{2.2}{
    d\expct{X} \arreq \frac{1}{m}\expct{P}\,dt + \sqrt{8\eta k} \Vx\,dW\\
    d\expct{P} \arreq -m\omega_0^{\,2}\expct{X}\,dt + \sqrt{8\eta k} \Cxp\,dW\\
    \partial_t \Vx \arreq \frac{2}{m} \Cxp - 8\eta k \Vx^{\,2}\\
    \partial_t \Vp \arreq -2m\omega_0^{\,2} \Cxp +2\hbar^2 k- 8\eta k \Cxp^{\,2}\\
    \partial_t \Cxp \arreq \frac{1}{m} \Vp -m\omega_0^{\,2}\Vx-
       8\eta k \Vx\Cxp.
  }
  \label{xmeasmoments}
\end{equation}
Notice that in the variance equations, the $dW$ terms vanished, due to the
assumption of a Gaussian state, which implies the following relations for the 
moments \cite{Habib04}:
\begin{equation}
  \eqnarr{1.5}{
    \expct{X^3} \arreq \expct{X}^3+3\expct{X}\Vx\\
    \ds\frac{1}{2}\expct{[X,P^2]_+} \arreq \expct{X}\expct{P}^2+2\expct{P}\Cxp+\expct{X}\Vp\\
    \ds\frac{1}{2}\expct{[X,[X,P]_+]_+} \arreq \expct{X}\expct{P}^2+2\expct{X}\Cxp+\expct{P}\Vx.
  }
\end{equation}
For the reader wishing to become better acquainted with 
continuous measurement theory,
the derivation of Eqs.~(\ref{xmeasmoments}) is an excellent exercise.
The derivation is straightforward, the only subtlety being the
second-order It\^o terms in the variances.  For example, the 
equation of motion for the position variance starts as
\begin{equation}
  d\Vx = d\expct{X^2}-2\expct{X} d\expct{X} - (d\expct{X})^2.
  \label{vxexample}
\end{equation}
The last, quadratic term is important in producing the effect that the
measured quantity becomes more certain.

In examining Eqs.~(\ref{xmeasmoments}), we can simply use the coefficients
to identify the source and thus the interpretation of each term.
The first term in each equation is due to the natural 
Hamiltonian evolution of the harmonic oscillator.
Terms originating from the $\mathcal{D}[c]\rho$ component are proportional
to $k\,dt$ but not $\eta$; in fact, the only manifestation of this term
is the $\hbar^2 k$ term in the equation of motion for $\Vp$.
Thus, a position measurement with rate constant $k$ produces momentum
diffusion (heating) at a rate $\hbar^2 k$, as is required to maintain the
uncertainty principle as the position uncertainty contracts due 
to the measurement.

There are more terms here originating from the $\mathcal{H}[c]\rho$ component
of the master equation, and they are identifiable since they
are proportional to either $\sqrt{\eta k}$ or $\eta k$.  
The $dW$ terms in the equations for $\expct{X}$ and $\expct{P}$ represent
the stochastic nature of the position measurement.  That is, during each
small time interval, the wave function collapses slightly, but we don't know 
exactly where it collapses to.  This stochastic behavior is
precisely the same behavior that we saw in Eq.~(\ref{alphastochastic}).
The more subtle point here lies with the nonstochastic terms proportional
to $\eta k$, which came from the second-order term [for example,
in Eq.~(\ref{vxexample})] where It\^o calculus generates a 
nonstochastic term from $dW^2=dt$.  Notice in particular the term of
this form in the $\Vx$ equation, which acts as a \textit{damping} term
for $\Vx$.  This term represents the certainty gained via the measurement
process.  The other similar terms are less clear in their interpretation,
but they are necessary to maintain consistency
of the evolution.

Note that we have made the assumption of a Gaussian initial state
in deriving these equations, but this assumption is not very restrictive.
Due to the linear potential and the Gaussian POVM for the measurement
collapse, these equations of motion preserve the Gaussian form of the initial
state.  The Gaussian POVM additionally converts arbitrary initial states into 
Gaussian states at long times.
Furthermore, the assumption of a Gaussian POVM is not 
restrictive---under the assumption of sufficiently high noise bandwidth,
the central-limit theorem guarantees that temporal
coarse-graining yields Gaussian noise for \textit{any} 
POVM giving random deviates with bounded variance.

\subsection{Dissipation}

The position measurement above is an example of a Hermitian measurement operator.
But what happens when the measurement operator is antihermitian?
As an example, we will consider the annihilation operator for the harmonic oscillator
by setting
$c=\sqrt{\gamma}\,a$,
where 
\begin{equation}
  a = \frac{1}{\sqrt{2}x_0}X+i\frac{x_0 }{\sqrt{2}\hbar}P
\end{equation}
and
\begin{equation}
  x_0 := \sqrt{\frac{\hbar}{m\omega_0}}.
\end{equation}
The harmonic oscillator with this type of measurement models, for example,
the field of an optical cavity whose output is monitored via
\textit{homodyne detection}, 
where the cavity output is mixed on a beamsplitter 
with another optical field.  (Technically, in homodyne detection, the
field must be the same as the field driving the cavity; mixing with other fields
corresponds to \textit{heterodyne} detection.)
A procedure very similar to the one above gives the following cumulant equations for the
conditioned evolution in this case:
\begin{equation}
  \eqnarr{2.2}{
    d\expct{X} \arreq \frac{1}{m}\expct{P}\,dt -\frac{\gamma}{2}\expct{X}\,dt \\ &&\ds 
         {}+\sqrt{2\eta \gamma\frac{m\omega_0}{\hbar}}\left( \Vx-\frac{\hbar}{2m\omega_0}\right) dW\\
    d\expct{P} \arreq -m\omega_0^{\,2}\expct{X}\,dt -\frac{\gamma}{2}\expct{P}\,dt \\ &&\ds 
         {}+\sqrt{2\eta \gamma\frac{m\omega_0}{\hbar}} \Cxp\,dW\\
    \partial_t \Vx \arreq \frac{2}{m} \Cxp -\gamma\left(\Vx-\frac{\hbar}{2m\omega_0}\right)\\ &&\ds
         {}-2\eta\gamma\frac{m\omega_0}{\hbar}\left(\Vx-\frac{\hbar}{2m\omega_0}\right)^2 \\ 
    \partial_t \Vp \arreq -m\omega_0^{\,2} \Cxp 
            -\gamma\left(\Vp-\frac{m\omega_0\hbar}{2}\right) \\ &&\ds
            {}-2\eta\gamma\frac{m\omega_0}{\hbar} \Cxp^{\,2}
            \\ 
    \partial_t \Cxp \arreq \frac{1}{m} \Vp -m\omega_0^{\,2}\Vx
        -\gamma \Cxp \\&&\ds
       {}-2\eta\gamma\frac{m\omega_0}{\hbar} \Cxp\left(\Vx-\frac{\hbar}{2m\omega_0}\right) . 
  }
\end{equation}
The moment equations seem more complex in this case, but are still
fairly simple to interpret.

First, consider the unconditioned evolution of the 
means $\expct{X}$ and $\expct{P}$, where we average over all possible
noise realizations.  
Again, since $\dexpct{\rho\, dW}=0$, we can simply set $dW=0$ in the above
equations, and we will drop the double angle brackets for brevity.
The Hamiltonian evolution terms
are of course the same, but now we see extra damping terms.
Decoupling these two equations 
gives an equation of the usual form for the damped harmonic oscillator for the 
mean position:
\begin{equation}
  \partial_t^{\,2}{\expct{X}} + \gamma\partial_t{\expct{X}} 
   + \left(\omega_0^{\,2}+\frac{\gamma^2}{4}\right)\expct{X} = 0.
\end{equation}
Note that we identify the frequency $\omega_0$ 
here as the actual oscillation frequency $\omega_\gamma$ of the damped oscillator,
given by $\omega_\gamma^{\,2} = \omega_0^{\,2} - \gamma^2/4$, and
not the resonance frequency $\omega_0$ that appears the usual form of the
classical formula.

The noise terms in these equations correspond to \textit{nonstationary
diffusion}, or diffusion where the transport rate depends on the
state of the system.  Note that under such a diffusive process,
the system will tend to come to rest in configurations where 
the diffusion coefficient vanishes, an effect closely related to
the ``blowtorch theorem'' \cite{Landauer93}. 
Here, this corresponds to $\Vx=\hbar/2m\omega_0$ and $\Cxp=0$.

The variance equations also contain unconditioned damping
terms (proportional to $\gamma$ but not $\eta$).
These damping terms cause the system to equilibrate with the same
variance values as noted above; they also produce the extra
equilibrium value
$\Vp=m\omega_0 \hbar/2$.
The conditioning terms (proportional to $\eta$) merely accelerate the
settling to the equilibrium values.
Thus, we see that the essential effect of the antihermitian 
measurement operator is to damp the energy from the system,
whether it is stored in the centroids or in the variances.
In fact, what we see is that this measurement process
selects \textit{coherent states}, states that have the same
shape as the harmonic-oscillator ground state, but whose centroids
oscillate along the classical harmonic-oscillator trajectories.

\section{Physical Model of a Continuous Measurement: Atomic Spontaneous
Emission}

To better understand the nature of continuous measurements, we will
now consider in detail an example of how a continuous measurement
of position
arises in a fundamental physical system: a single atom interacting
with light.
Again, to obtain weak measurements, we do not make projective measurements
directly on the atom, but rather we allow the atom to become entangled
with an auxiliary quantum system---in this case, the electromagnetic field---and
then make projective measurements on the auxiliary system (in this case,
using a photodetector).  It turns out that this one level of separation
between the system and the projective measurement is the key
to the structure of the formalism.  Adding more elements to the chain
of quantum-measurement devices does not change the fundamental structure that
we present here.

\subsection{Master Equation for Spontaneous Emission}

We begin by considering the interaction of the atom with the electromagnetic
field.  In particular, treating the field quantum mechanically allows
us to treat spontaneous emission.  These spontaneously emitted photons
can then be detected to yield information about the atom.

\subsubsection{Decay of the Excited State}

We will give a brief treatment following the approach of 
Weisskopf and Wigner~\cite{Weisskopf30, Scully97, Milonni94}.
Without going into detail about the quantization of the electromagnetic
field, we will simply note that the quantum description of the 
field involves associating a quantum harmonic oscillator with each field mode 
(say, each plane wave of a particular wave vector $\mathbf{k}$ and
definite polarization).  Then for a two-level atom with 
ground and excited levels $\gket$ and $\eket$, respectively, the
uncoupled Hamiltonian for the atom and a \textit{single} field mode
is
\begin{equation}
  H_0 = \hbar\omega_{0}\sigma^\dagger\sigma
       +\hbar\omega
     \left(a^\dagger a+\frac{1}{2}\right).
\end{equation}
Here, $\omega_0$ is the transition frequency of the atom, 
$\omega$ is the frequency of the field mode,
$\sigma:=\gket\ebra$ is the atomic lowering operator (so that
$\sigma^\dagger\sigma=\eket\ebra$ is the excited-state projector), and
$a$ is the field (harmonic oscillator) annihilation operator.
The interaction between the atom and field is given in the dipole
and rotating-wave approximations by the interaction Hamiltonian
\begin{equation}
  \HAF =\hbar\left(g\sigma^\dagger a+g^*\sigma a^\dagger\right),
  \label{HAF}
\end{equation}
where $g$ is a coupling constant that includes the volume of the mode,
the field frequency, and the atomic dipole moment.
The two terms here are the ``energy-conserving'' processes
corresponding to photon absorption and emission.

In the absence of externally applied fields,
we can write the state vector as the superposition of the states
\begin{equation}
  \psiket = \ce \eket + \cg\goneket,
  \label{superpos}
\end{equation}
where the uncoupled eigenstate $\ket{\alpha, n}$ denotes the atomic state
$\ket\alpha$ and the $n$-photon field state, and the omitted photon number
denotes the vacuum state:
$\ket{\alpha}\equiv\ket{\alpha, 0}$.
These states form an effectively complete basis,
since no other states are coupled to these by
the interaction (\ref{HAF}).
We will also assume that the atom is initially excited, so that
$\ce(0)=1$ and $\cg(0)=0$.

The evolution is given by the Schr\"odinger equation,
\begin{equation}
  \partial_t\psiket = -\,\frac{i}{\hbar}(H_0+\HAF)\psiket,
\end{equation}
which gives, upon substitution of (\ref{superpos}) and dropping the
vacuum energy offset of the field,
\begin{equation}
  \eqnarr{1.0}{
   \partial_t\ce\arreq -i\omega_0\ce-i g\cg\\
   \partial_t\cg\arreq -i\omega\cg-i g^*\ce.
  }
\end{equation}
Defining the slowly varying amplitudes $\cet:=\ce e^{i\omega_0t}$ and
$\cgt:=\cg e^{i\omega t}$, we can rewrite these as
\begin{equation}
  \eqnarr{1.0}{
   \partial_t\cet\arreq -i g\cgt e^{-i(\omega-\omega_0)t}\\
   \partial_t\cgt\arreq -i g^*\cet e^{i(\omega-\omega_0)t}.
  }
  \label{slowlyvaryingeqs}
\end{equation}
To decouple these equations, we first integrate the equation for $\cgt$:
\begin{equation}
   \cgt(t)= -i g^*\int_0^t dt'\,\cet(t') e^{i(\omega-\omega_0)t'}.
\end{equation}
Substituting this into the equation for $\cet$,
\begin{equation}
   \partial_t\cet= - |g|^2 \int_0^t dt'\, \cet(t') e^{-i(\omega-\omega_0)(t-t')},
\end{equation}
which gives the evolution for the excited state coupled to a single field mode.

Now we need to sum over all field modes.  In free space, we can
integrate over all possible plane waves, labeled by the wave vector
$\mathbf{k}$ and the two possible polarizations $\zeta$ for each wave vector.
Each mode has a different frequency $\omegak=ck$, and we must
expand the basis so that a photon can be emitted into any mode:
\begin{equation}
  \psiket = \ce \eket + \sum_{\mathbf{k},\zeta} \ckz \gkzket.
\end{equation}
Putting in the proper
form of the coupling constants $g_\mathbf{k}$ for each mode in the free-space
limit, it turns out that the equation of motion becomes
\begin{equation}
  \begin{array}{l}
    \ds\partial_t\cet
    =\\ \ds-\,
    \frac{d_\mathrm{ge}^2}{6\epsilon_0\hbar (2\pi)^3}
    \sum_\zeta \int d\mathbf{k}\, \omegak
    \ds\int_0^t dt'\,\cet(t') e^{-i(\omegak-\omega_0)(t-t')},
  \end{array}
\end{equation}
where $\mathbf{d}_\mathrm{ge}:=\gbra\mathbf{d}\eket$ is the dipole matrix
element characterizing the atomic transition strength.  The polarization sum
simply contributes a factor of 2, while carrying out the angular integration
in spherical coordinates gives
\begin{equation}
    \partial_t\cet 
    = -\,
    \frac{d_\mathrm{ge}^2}{6\pi^2\epsilon_0\hbar c^3}
     \int_0^\infty \!\!d\omega\,\omega^3
    \int_0^t dt'\,\cet(t') e^{-i(\omegak-\omega_0)(t-t')}.
  \label{WWlastbeforeeval}
\end{equation}

We can now note that $\cet(t')$ varies slowly on optical time scales.
Also, $\omega^3$ is slowly varying compared to the exponential factor
in Eq.~(\ref{WWlastbeforeeval}), which oscillates rapidly (at least for 
large times $t$) about zero
except when $t\approx t'$ and $\omega\approx\omega_0$.  Thus, we will
get a negligible contribution from the $\omega$ integral away from $\omega=\omega_0$.
We will therefore make the replacement $\omega^3 \longrightarrow \omega_0^{\,3}$:
\begin{equation}
  \eqnarr{2.2}{
    \partial_t\cet 
    \arreq -
    \frac{\omega_0^{\,3} d_\mathrm{ge}^2}{6\pi^2\epsilon_0\hbar c^3}
    \int_0^\infty d\omega\,  
    \int_0^t dt'\,\cet(t') e^{-i(\omegak-\omega_0)(t-t')}.
  }
  \label{WWlastbeforeeval2}
\end{equation}
The same argument gives
\begin{equation}
  \eqnarr{1.2}{
  \ds\int_0^\infty d\omega\, e^{-i(\omegak-\omega_0)(t-t')} &{}\approx{}&\ds
  \int_{-\infty}^\infty d\omega\, e^{-i(\omegak-\omega_0)(t-t')}\\ \arreq
  2\pi\delta(t-t').
  }
\end{equation}
We can see from this that our argument here about the exponential factor is equivalent to the Markovian
approximation, where we assume that the time derivative of the 
quantum state depends \text{only} on the state at the present time. Thus,
\begin{equation}
  \eqnarr{2.2}{
    \partial_t\cet 
    \arreq -
    \frac{\omega_0^{\,3} d_\mathrm{ge}^2}{3\pi\epsilon_0\hbar c^3}
    \int_0^t dt'\,\cet(t') \delta (t-t')\\
    \arreq -
    \frac{\omega_0^{\,3} d_\mathrm{ge}^2}{3\pi\epsilon_0\hbar c^3}
    \frac{\cet(t)}{2}.\\
  }
  \label{WWlastbeforeeval3}
\end{equation}
Here, we have split the $\delta$-function since the upper limit of the $t'$ integral
was $t$, in view of the original form (\ref{WWlastbeforeeval2}) for the $t'$ integral,
where the integration limit is centered at the peak of the exponential factor.
We can rewrite the final result as
\begin{equation}
  \partial_t\cet = -\,\frac{\Gamma}{2}\cet,
  \label{cedecay}
\end{equation}
where the spontaneous decay rate is given by
\begin{equation}
  \Gamma := \frac{\omega_0^{\,3} d^2_\mathrm{ge}}{3\pi\epsilon_0\hbar c^3}.
  \label{gammaresult}
\end{equation}
This decay rate is of course defined so that the \textit{probability} decays 
exponentially at 
the rate $\Gamma$.
Also, note that
\begin{equation}
  \partial_t\ce = \left(-i\omega_0-\frac{\Gamma}{2}\right)\ce
\end{equation}
after transforming out of the slow variables.

\subsubsection{Form of the Master Equation}

We now want to consider the \textit{reduced}
density operator for the evolution of the atomic state,
tracing over the state of the field.
Here we will compute the individual matrix elements 
\begin{equation}
  \rho_{\alpha\beta}:=\expct{\alpha|\rho|\beta}
\end{equation}
for the atomic state.

The easiest matrix element to treat is the excited-level population,
\begin{equation}
  \rho_{\mathrm{ee}} = \ce\ce^*.
  \label{rhoeedef}
\end{equation}
Differentiating this equation and using (\ref{cedecay}) gives
\begin{equation}
  \partial_t \rho_\mathrm{ee} = -\Gamma \rho_\mathrm{ee}.
  \label{rhoeedecay}
\end{equation}
The matrix element for the ground-state population follows from summing
over all the other states:
\begin{equation}
  \rho_\mathrm{gg} :=  \sum_{\zeta}\int d\mathrm{k}\, \ckzt\ckzt^*.
  \label{rhoggmast}
\end{equation}
Notice that the states $\eket$ and $\gket$ are effectively degenerate,
but when we eliminate the field, we want $\eket$ to have $\hbar\omega_0$
more energy than the ground state.  The shortcut for doing this
is to realize that the latter situation corresponds to the 
``interaction picture'' with respect to the field, where
we use the slowly varying ground-state
amplitudes $\ckzt$ but the standard excited-state amplitude $\ce$.
This explains why we use regular coefficients in
Eq.~(\ref{rhoeedef})
but the slow variables in
Eq.~(\ref{rhoggmast}).
Since by construction $\rho_\mathrm{ee} + \rho_\mathrm{gg} = 1$,
\begin{equation}
  \partial_t \rho_\mathrm{gg} = \Gamma \rho_\mathrm{ee}.
\end{equation}
Finally, the coherences are
\begin{equation}
  \rho_\mathrm{ge} := \sum_\zeta\int d\mathbf{k}\,\ckzt \ce^*,
  \hspace{5mm}\rho_\mathrm{eg}=\rho_\mathrm{ge}^*,
\end{equation}
and so the corresponding equation of motion is
\begin{equation}
  \partial_t\rho_\mathrm{ge} = \sum_\zeta\int d\mathbf{k}\,\ckzt 
   \left(i\omega_0 -\frac{\Gamma}{2}\right)\ce^*
   = \left(i\omega_0 -\frac{\Gamma}{2}\right)\rho_\mathrm{ge}.
\end{equation}
We have taken the time derivatives of the $\ckzt$ to be zero here. From 
Eq.~(\ref{slowlyvaryingeqs}), the time derivatives, when summed over all
modes, will in general correspond to a sum over amplitudes with
rapidly varying phases, and thus their contributions will cancel.

Notice that what we have derived are exactly the same matrix elements generated
by the master equation
\begin{equation}
  \partial_t\rho = -\,\frac{i}{\hbar}[\HA,\rho] +\Gamma\mathcal{D}[\sigma]\rho,
  \label{SEmasteqn}
\end{equation}
where the form of $\mathcal{D}[\sigma]\rho$ is given by Eq.~(\ref{lindbladsuperop}),
and the atomic Hamiltonian is
\begin{equation}
  \HA := \hbar\omega_0 \ket{\mathrm{e}}\bra{\mathrm{e}}.
\end{equation}
That is, the damping term here represents the same damping
as in the optical Bloch equations.

\subsection{Photodetection: Quantum Jumps and the Poisson Process}\label{section:photodetection}

In deriving Eq.~(\ref{SEmasteqn}), we have ignored the
state of the field.  Now we will consider what happens when we measure
it.  In particular, we will assume that we make \textit{projective}
measurements of the field photon number in every mode, not distinguishing
between photons in different modes.
It is this extra interaction that will yield the continuous measurement
of the atomic state.

\relax From Eq.~(\ref{rhoeedecay}), the transition probability in a time
interval of length $dt$ is $\Gamma\rho_\mathrm{ee}\,dt=\Gamma\expct{\sigma^\dagger\sigma}dt$,
where we recall that $\sigma^\dagger\sigma =\ket{\mathrm{e}}\bra{\mathrm{e}}$ is the
excited-state projection operator.
Then assuming an ideal detector that detects photons at all frequencies,
polarizations, and angles, there are two possibilities during this time interval:
\begin{enumerate}
\item \textbf{No photon detected.} The detector does not ``click'' in this
case, and this possibility happens with probability 
$1-\Gamma\expct{\sigma^\dagger\sigma}dt$.  The same construction 
as above for the master equation carries through, so we keep
the equations of motion for $\rho_\mathrm{ee}$, $\rho_\mathrm{eg}$,
and $\rho_\mathrm{ge}$.  However, we do not keep the same equation
for $\rho_\mathrm{gg}$: no photodetection implies that the atom
does not return to the ground state.  Thus, $\partial_t \rho_\mathrm{gg}=0$.
This case is thus generated by the master equation
\begin{equation}
  \partial_t\rho =-\,\frac{i}{\hbar}[\HA,\rho]-\frac{\Gamma}{2}[\sigma^\dagger\sigma,\rho]_+ .
\end{equation}
This evolution is unnormalized since $\mathrm{Tr}[\rho]$ decays to zero at long times.
We can remedy this by explicitly renormalizing the state $\rho(t+dt)$,
which amounts to adding one 
term to the master equation, as in Eq.~(\ref{gensme1}):
\begin{equation}
  \partial_t\rho =-\,\frac{i}{\hbar}[\HA,\rho]-\frac{\Gamma}{2}[\sigma^\dagger\sigma,\rho]_+ 
    +\Gamma\expct{\sigma^\dagger\sigma}\rho.
\end{equation}

\item \textbf{Photon detected.}  A click on the photodetector
occurs with probability 
$\Gamma\expct{\sigma^\dagger\sigma}dt$.
The interaction Hamiltonian $\HAF$ contains a term of the form $\sigma a^\dagger$,
which tells us that photon creation (and subsequent detection) 
is accompanied by lowering of the atomic state.
Thus, the evolution for this time interval is given by the reduction
\begin{equation}
  \rho(t+dt) = \frac{\sigma\rho(t)\sigma^\dagger}{\expct{\sigma^\dagger\sigma}}.
\end{equation}
We can write this in differential form as
\begin{equation}
  d\rho = \frac{\sigma\rho\sigma^\dagger}{\expct{\sigma^\dagger\sigma}}-\rho.
\end{equation}

\end{enumerate}
The overall evolution is stochastic, with either case occurring during a
time interval $dt$ with the stated probabilities.

We can explicitly combine these two probabilities by defining
a stochastic variable $dN$, called the \textit{Poisson process}.
In any given time interval $dt$, $dN$ is unity with probability
$\Gamma\expct{\sigma^\dagger\sigma}dt$ and zero otherwise.
Thus, we can write the average over all possible stochastic histories
as
\begin{equation}
  \dexpct{dN} = \Gamma\expct{\sigma^\dagger\sigma}dt.
\end{equation}
Also, since $dN$ is either zero or one, the process satisfies $dN^2=dN$.
These last two features are sufficient to fully characterize the 
Poisson process.

\begin{figure}[tb]
  \begin{center}
     \includegraphics[width=1.0\hsize]{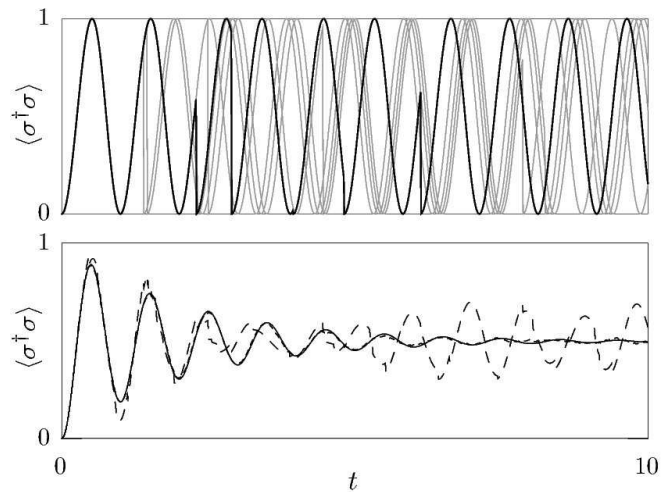}
   \end{center}
  \vspace{-5mm}
  \caption
        {Quantum jumps in a driven two-level atom.
         Top: evolution of the excited-state probability
         for a single atom (quantum trajectory) with jumps to the ground state,
         corresponding to a detected photon.
         Four other trajectories are included to illustrate the
         dephasing due to the
         random nature of the jumps.
         Bottom: ensemble-averaged 
         excited-state probability computed from the 
         master equation (solid line), an
         average of 20 trajectories (dashed line), and an
         average of 2000 trajectories (dotted line).
         Time is measured in units of $2\pi/\Omega$ [see 
         Eq.~(\ref{drivingHam})],
         and the decay rate is $\Gamma=0.1$ in the same units.
	\label{fig:jumps}}
\end{figure}

Now we can add the two above possible cases together, with a weighting factor
of $dN$ for the second case:
\begin{equation}
  \eqnarr{2.1}{
  d\rho \arreq-\,\frac{i}{\hbar}[\HA,\rho]dt-\frac{\Gamma}{2}[\sigma^\dagger\sigma,\rho]_+dt 
    +\Gamma\expct{\sigma^\dagger\sigma}\rho\,dt\\ &&\ds
  {}+\left( \frac{\sigma\rho\sigma^\dagger}{\expct{\sigma^\dagger\sigma}}-\rho\right)dN.
  }
  \label{jumpmasteqn}
\end{equation}
It is unnecessary to include a weighting factor
of $(1-dN)$ for the first term,
since $dN\,dt=0$.
It is easy to verify that this master equation is equivalent to
the stochastic Schr\"odinger equation
\begin{equation}
  \eqnarr{2.1}{
  d\ket\psi \arreq -\,\frac{i}{\hbar} \HA\ket\psi dt
   + \frac{\Gamma}{2}\left(\expct{\sigma^\dagger\sigma}-\sigma^\dagger\sigma\right)\ket\psi dt \\&&\ds
  {}+\left( \frac{\sigma}{\sqrt{\expct{\sigma^\dagger\sigma}}}-1\right)\ket\psi\, dN,
  }
  \label{jumpsse}
\end{equation}
again keeping terms to second order and using $dN^2=dN$.
Stochastic Schr\"odinger equations of this form are popular for simulating
master equations, since if the state vector has $O(n)$ components, the
density matrix will have $O(n^2)$ components, and thus is much more
computationally expensive to solve.  If $s$ solutions
(``quantum trajectories'') of the stochastic Schr\"odinger equation
can be averaged together to obtain a sufficiently accurate
solution to the master equation and $s\ll n$, then this
Monte-Carlo-type method is computationally
efficient for solving the master equation.
This idea is illustrated in Fig.~\ref{fig:jumps},
which shows quantum trajectories for the two-level atom driven 
by a field according to the Hamiltonian (\ref{drivingHam})
in Section~\ref{section:quasiequilibrium}.
As many trajectories are averaged together, the 
average converges to the master-equation solution for the ensemble average.
(About 20,000 trajectories are necessary for the Monte-Carlo average
to be visually indistinguishable from the master-equation solution on
the time scale plotted here.)
Note that the ``Rabi oscillations'' apparent here are distorted slightly
by the nonlinear renormalization term in Eq.~(\ref{jumpsse})
from the usual sinusoidal oscillations in the absence of spontaneous emission.
However, the damping rate in Fig.~\ref{fig:jumps} is small, so the distortion
is not visually apparent.
``Unravellings'' \cite{Carm93} of this form are much easier to solve computationally than
``quantum-state diffusion'' unravellings involving $dW$.
Of course, it is important for more than just a numerical method, since
this gives us a powerful formalism for handling photodetection.

To handle the case of photodetectors with less than ideal efficiency $\eta$,
we simply combine the conditioned and unconditioned stochastic master
equations, with weights $\eta$ and $1-\eta$, respectively:
\begin{equation}
  \eqnarr{2.1}{
  d\rho \arreq
    -\,\frac{i}{\hbar}[\HA,\rho]dt
    +\eta\frac{\Gamma}{2}\left[\expct{\sigma^\dagger\sigma}-\sigma^\dagger\sigma,\rho\right]_+dt
    \\ &&\ds {}+(1-\eta)\Gamma\mathcal{D}[\sigma]\rho\,dt 
  +\left( \frac{\sigma\rho\sigma^\dagger}{\expct{\sigma^\dagger\sigma}}-\rho\right)dN\\
    \arreq-\,\frac{i}{\hbar}[\HA,\rho]dt+\Gamma\mathcal{D}[\sigma]\rho\,dt
    +\eta\Gamma\expct{\sigma^\dagger\sigma}\rho\,dt\\ &&\ds
    {} -\eta\Gamma \sigma\rho\sigma^\dagger\,dt
  +\left( \frac{\sigma\rho\sigma^\dagger}{\expct{\sigma^\dagger\sigma}}-\rho\right)dN.
  }
  \label{partialefficiencyjumpmasteq}
\end{equation}
The Poisson process is modified in this case such that
\begin{equation}
  \dexpct{dN} = \eta\Gamma\expct{\sigma^\dagger\sigma}dt
\end{equation}
to account for the fact that fewer photons are detected.

\subsection{Imaged Detection of Fluorescence}

\subsubsection{Center-of-Mass Dynamics}
Now we want to consider how the evolution of the atomic internal state
influences the atomic center-of-mass motion.
To account for the external atomic motion, we use the center-of-mass
Hamiltonian
\begin{equation}
  \HCM = \frac{p^2}{2m} + V(x)
\end{equation}
in addition to the internal atomic Hamiltonian $\HA$.
We also need to explicitly include the spatial dependence of the field
by letting
\begin{equation}
  g_\mathbf{k}\longrightarrow g_\mathbf{k} e^{i\mathbf{k}\cdot\mathbf{r}}
\end{equation}
in the interaction Hamiltonian (\ref{HAF}).
In the weak-excitation limit, we can take $\mathbf{k}$ to
have the value $\kLbf$ of an externally applied probe field 
(the emitted photons are elastically scattered from the incident field).

To include the center of mass in the atomic state, we can explicitly write the
state in terms of 
momentum-dependent coefficients as
\begin{equation}
  \ket\psi = \int d\mathbf{p}\,\psi_\mathrm{e}(\mathbf{p})\ket{\mathbf{p},\mathrm{e}}
   + \sum_{\mathbf{k},\zeta} \psi_{\mathbf{k},\zeta}(\mathbf{p})\ket{\mathbf{p},\mathrm{g},1_{\mathbf{k},\zeta}}.
\end{equation}
Notice that the new interaction Hamiltonian
\begin{equation}
  \HAF =\sum_{\mathbf{k},\zeta}
      \hbar\left(g_{\mathbf{k},\zeta}a_{\mathbf{k},\zeta}\sigma^\dagger e^{i\mathbf{k}\cdot\mathbf{r}}
          +g^*_{\mathbf{k},\zeta}a^\dagger_{\mathbf{k},\zeta}\sigma e^{-i\mathbf{k}\cdot\mathbf{r}}\right)
  \label{HAFp}
\end{equation}
couples the state $\ket{\mathbf{p},\mathrm{e}}$ 
to the states $\ket{\mathbf{p}-\hbar\mathbf{k},\mathrm{g},1_{\mathbf{k},\zeta}}$
(in the momentum basis),
giving rise to the atomic momentum recoil from spontaneous emission.
(The additional recoil due to the absorption of the photon comes about
by examining the coupling to the driving field.)
The derivation of the last section carries through here with the
replacement 
\begin{equation}
  \sigma\longrightarrow\sigma e^{-i\kLbf\cdot\mathbf{r}}.
\end{equation}
Summing over all possible emission directions, the
unconditioned master equation (\ref{SEmasteqn}) becomes
\begin{equation}
  \partial_t\rho = -\,\frac{i}{\hbar}[\HA+\HCM,\rho] 
  +\Gamma\int d\Omega\, f(\theta,\phi)\, \mathcal{D}\!\left[\sigma e^{-i\kLbf\cdot\mathbf{r}}\right]\rho,
  \label{SEmasteqnwithp}
\end{equation}
where $f(\theta,\phi)$ is the normalized classical angular distribution
for the radiated light, which here represents the angular probability
distribution for the emitted photons.

Applying the same reasoning here as for the quantum-jump master equation
(\ref{jumpmasteqn}), we obtain
\begin{equation}
  \eqnarr{2.1}{
  d\rho \arreq-\,\frac{i}{\hbar}[\HA+\HCM,\rho]dt
    +\frac{\Gamma}{2}\left[\expct{\sigma^\dagger\sigma}-\sigma^\dagger\sigma,\rho\right]_+dt 
   \\ &&\ds
  {}+
   \int d\Omega
    \left( \frac{\sigma e^{-i\mathbf{k}\cdot\mathbf{r}}\rho\sigma^\dagger e^{i\mathbf{k}\cdot\mathbf{r}}}{\expct{\sigma^\dagger\sigma}}-\rho\right) \frac{dN(\theta,\phi)}{d\Omega},
  }
  \label{jumpmasteqnwithp}
\end{equation}
where 
\begin{equation}
  \dexpct{\frac{dN(\theta,\phi)}{d\Omega} }
  = \Gamma\expct{\sigma^\dagger\sigma} f(\theta,\phi)\, dt
\end{equation}
as before.  
We can simplify this equation by carrying out the angular integral,
defining $dN$ to be one whenever $\max[dN(\theta,\phi)]=1$.
The result is
\begin{equation}
  \eqnarr{2.1}{
  d\rho \arreq-\,\frac{i}{\hbar}[\HA+\HCM,\rho]dt
    +\frac{\Gamma}{2}\left[\expct{\sigma^\dagger\sigma}-\sigma^\dagger\sigma,\rho\right]_+dt 
   \\ &&\ds
  {}+
    \left( \frac{\sigma e^{-i\kLbf\cdot\mathbf{r}}\rho\sigma^\dagger e^{i\mathbf{k}\cdot\mathbf{r}}}{\expct{\sigma^\dagger\sigma}}-\rho\right) dN
  }
  \label{jumpmasteqnwithpnoang}
\end{equation}
with
\begin{equation}
  \dexpct{dN}
  = \Gamma\expct{\sigma^\dagger\sigma}  dt
\end{equation}
as before.
The angles $\theta$ and $\phi$ are then stochastic variables
with probability density $f(\theta,\phi)\sin\theta$.

\subsubsection{Imaging}

The above master equation (\ref{jumpmasteqnwithp})
is for an \textit{angle-resolving} detector.  What we see is that angle-resolved
detection keeps explicit track of the atomic momentum kicks due to 
spontaneous emission.
An \textit{imaging} detector, on the other hand, gives up resolution
of the direction of the emitted photon wave vector $\mathbf{k}$,
thus obtaining instead some position information about the atom.
An imaging system operates by summing fields from many directions together
and then detecting the resulting interference pattern.
The procedure for obtaining the measurement operators for
the imaging system is as follows \cite{Holland96, Greenwood97}.
Notice that we can regard the master equation~(\ref{jumpmasteqnwithp})
as a normal jump process of the form
(\ref{jumpmasteqn}), with measurement operators
\begin{equation}
  \sigma(\theta,\phi) = \sqrt{f(\theta,\phi)}\,\sigma e^{i\kL z\cos\theta},
\end{equation}
where we sum over all possible emission angles.  In writing down
this operator, we are specializing to one-dimensional motion along
the $z$-axis ($x=y=0$), so we only require the $z$-component 
$k\cos\theta$ of $\mathbf{k}$.
This operator ranges from $-1$ to $1$ in $\cos\theta$ and from
$0$ to $2\pi$ in $\phi$.
Thus, we can write down Fourier coefficients, since these
functions are defined on a bounded domain, with two indices
$\alpha$ and $\beta$:
\begin{widetext}
\begin{equation}
  \tilde \sigma_{\alpha\beta} = 
    \frac{\sigma}{\sqrt{4\pi}} \int_0^{2\pi} d\phi \int_{-1}^1
     d(\cos\theta)\,\sqrt{f(\theta,\phi)}\, e^{i\kL z\cos\theta} 
     e^{-i\alpha\pi\cos\theta} e^{-i\beta\phi}.
  \label{sigmaab}
\end{equation}
\end{widetext}
If we consider an atom whose radiation pattern is axially symmetric,
then performing the $\phi$ integral
amounts to letting $f(\theta,\phi)\longrightarrow f(\theta)/2\pi$,
since the integral is nonzero only for $\beta=0$.  Carrying this
out and suppressing the $\beta$ dependence,
\begin{equation}
  \tilde \sigma_{\alpha} = 
    \frac{\sigma}{\sqrt{2}}  \int_{-1}^1
     d(\cos\theta)\,\sqrt{f(\theta)}\, 
     e^{i\kL (z-\alpha\lambda/2)\cos\theta}.
  \label{sigmaab2}
\end{equation}
Notice that with the normalization convention for the Fourier coefficients
here,
\begin{equation}
  \int d\Omega\,\sigma^\dagger(\theta,\phi)\sigma(\theta,\phi)
   =\sum_\alpha \tilde\sigma^\dagger_\alpha \tilde \sigma_\alpha,
\end{equation}
so that the set of measurement operators is complete and properly normalized
in either basis.

Notice that the $\tilde\sigma_\alpha$ operators
contain localized functions of the position $z$, and thus correspond
to position measurements.
For example, a radiating atomic dipole oriented along the $z$-axis
has
\begin{equation}
  f(\theta) = \frac{3}{4}\sin^2\theta,
  \label{ftheta}
\end{equation}
which gives measurement operators of the form
\begin{equation}
  \tilde \sigma_{\alpha} = \sigma
    \sqrt{\frac{3\pi^2}{8}}
      \frac{J_1(\kL z_\alpha)}{\kL z_\alpha},
\end{equation}
where $z_\alpha := z-\alpha\lambda/2$, and $J_1(x)$ is an ordinary
Bessel function.
Notice also that the set of possible measurement values is not 
continuous, but rather is discretely spaced by $\lambda/2$.

\subsubsection{Gaussian Aperture}

\begin{figure}[tb]
  \begin{center}
    \includegraphics[width=0.5\hsize]{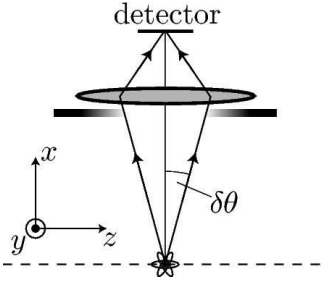}
  \end{center}
  \vspace{-5mm}
  \caption
        {Basic setup for imaging resonance fluorescence from
         a single atom as a continuous position measurement.
         Light scattered from a probe laser (not shown)
         is collected by a Gaussian aperture of angular half-width
         $\delta \theta$ and focused by a lens on a position-sensitive
         detector, such as a photodiode array.  The atom is
         constrained to move along the $z$-axis.
	\label{fig:imaging}}
\end{figure}

For the ideal imaging system we have considered here, 
the aperture extends over the
full $4\pi$ solid angle (requiring, for example, arbitrarily large
lenses on either side of the atom), 
though in practice it is rare to come anywhere
close to this extreme.
Thus, we will include the effects of an aperture that only allows the
imaging system to detect radiated light within a limited solid angle
(Fig.~\ref{fig:imaging}).
For mathematical convenience, we will choose an aperture with 
a Gaussian spatial profile.
We consider the above case of motion along the $z$-axis, with
the atomic dipole oriented along the $z$-axis.  Then photons
going into \textit{any} azimuthal angle $\phi$ are equivalent
as far as providing position information about the atom,
since the form of $\sigma(\theta,\phi)$ is independent of $\phi$.
Thus, it suffices to consider only the $\theta$ dependence of
the aperture, as any $\phi$ dependence contributes only by reducing 
the effective detection efficiency of the photodetector.
Intuitively, one expects a camera imaging system to be most effective
when oriented normal to the $z$-axis, so we choose the aperture
to be centered about $\theta=\pi/2$.
We thus take the \textit{intensity} transmission function of the aperture
to be
\begin{equation}
  T(\theta) = \exp\left[-\,\frac{2(\theta-\pi/2)^2}{(\delta\theta)^2}\right].
\end{equation}
The generalization of Eq.~(\ref{sigmaab2}) to this case is
\begin{equation}
  \tilde \sigma_{\alpha} = 
    \frac{\sigma}{\sqrt{2}}  \int_{-1}^1
     d(\cos\theta)\,\sqrt{T(\theta)f(\theta)}\, 
     e^{i\kL (z-\alpha\lambda/2)\cos\theta}.
\end{equation}
If $\delta\theta$ is small, then the integrand is only appreciable for
$\theta$ near $\pi/2$ due to the Gaussian factor.  Recentering the 
integrand, making the small-angle approximation 
in the rest of the integrand,
and extending the limits of integration, we find
\begin{equation}
  \eqnarr{2.3}{
  \tilde \sigma_{\alpha} \arreq
    \sigma\sqrt{\frac{3}{8}}  \int_{-\pi/2}^{\pi/2} \!\!\!\!\!
     d\theta\,\cos^2\theta \,
     e^{-i\kL (z-\alpha\lambda/2)\sin\theta}
     \exp\left[-\,\frac{\theta^2}{(\delta\theta)^2}\right]\\
    &{}\approx{}&\ds
    \sigma\sqrt{\frac{3}{8}}  \int_{-\infty}^{\infty} \!\!\!
     d\theta\,
     e^{-i\kL (z-\alpha\lambda/2)\theta}
     \exp\left[-\,\frac{\theta^2}{(\delta\theta)^2}\right]\\
    \arreq
    \sigma\sqrt{\frac{3\pi}{8}}
     \delta\theta\,
     \exp\left[-\left(\frac{\kL\,\delta\theta}{2}\right)^2\,\left(z-\frac{\alpha\lambda}{2}\right)^2\right]
     .
    }
\end{equation}
Thus, the measurement operator in this case is actually Gaussian.
We can write the fraction of photons transmitted by the aperture
as an efficiency
\begin{equation}
  \eta_\theta := \int_{-1}^1 d(\cos\theta)\, T(\theta)f(\theta)
     \approx \frac{3}{4}\sqrt{\frac{\pi}{2}}\,\delta\theta
\end{equation}
in the same regime of small $\delta\theta$.
Then the Gaussian measurement operators $\tilde\sigma_\alpha$
satisfy
\begin{equation}
  \sum_\alpha \tilde\sigma_\alpha^\dagger \tilde\sigma_\alpha
     =\eta_\theta \sigma^\dagger\sigma.
\end{equation}
This normalization is sensible, although as we will see later, 
$\eta_\theta$ turns out \textit{not} to be the
actual measurement efficiency.

\subsubsection{Spatial Continuum Approximation}

If an atom is initially completely delocalized, after one photon is
detected and the collapse operator $\tilde \sigma_{\alpha}$ applies,
the atom is reduced to a width of order
\begin{equation}
  \delta\alpha = \frac{1}{\kL\,\delta\theta} = \frac{\lambda}{2\pi\,\delta\theta}.
\end{equation}
Since this is much larger than the spacing
\begin{equation}
  \Delta\alpha = \frac{\pi}{\kL} = \frac{\lambda}{2},
\end{equation}
it is effectively impossible to ``see'' the discreteness of the measurement
record, and
it is a good approximation to replace the set of measurement operators
with a set corresponding to a continuous range of possible measurement
outcomes.
Since in the limit of small spacing $\Delta x$, it is a good approximation to
write an integral as a sum
\begin{equation}
  \sum_n  f(n\Delta x)  \,\Delta x
    = \int dx\, f(x)
\end{equation}
for an arbitrary function $f(x)$,
we can make the formal identification
\begin{equation}
  \tilde\sigma_\alpha \longrightarrow \frac{\tilde\sigma(\alpha)}{\sqrt{\Delta\alpha}}
\end{equation}
to obtain the continuum limit of the position collapse operators.
Thus, we have
\begin{equation}
  \eqnarr{2.3}{
  \tilde \sigma(\alpha) \arreq \int dz\ket{z}\bra{z}\,
    \sigma\sqrt{\eta_\theta}
      \frac{1}{\sqrt{\sqrt{2\pi}\,\delta\alpha}}
     \exp\left[-\,\frac{\left(z-\alpha\right)^2}{4(\delta\alpha)^2}\right]
     .
    }
\end{equation}
We have inserted the identity here to make this expression
a proper operator on the atomic center-of-mass state.
Again, $\alpha$ is now a continuous index with dimensions of length,
rather than an integer index.

Thus, from the form of Eq.~(\ref{partialefficiencyjumpmasteq}),
we can deduce the following form of the master equation for 
imaged photodetection through the Gaussian aperture:
\begin{equation}
  \eqnarr{2.1}{
  d\rho 
    \arreq-\,\frac{i}{\hbar}[\HA+\HCM,\rho]dt+\Gamma\int d\Omega\,\mathcal{D}[\sigma(\theta,\phi)]\rho\,dt \\&&\ds
    {}+\eta_\theta\Gamma\expct{\sigma^\dagger\sigma}\rho\,dt\\ &&\ds
    {} -\Gamma\int d\Omega\, T(\theta) \,\sigma(\theta,\phi)\,\rho\,\sigma^\dagger(\theta,\phi)\,dt \\&&\ds
  {}+\left[ \frac{\tilde\sigma(\alpha)\,\rho\,\tilde\sigma^\dagger(\alpha)}{\expct{\tilde\sigma^\dagger(\alpha)\tilde\sigma(\alpha)}}-\rho\right]dN.
  }
  \label{partialefficiencyjumpmasteqalpha}
\end{equation}
Recalling the normalization
\begin{equation}
  \int d\Omega\, T(\theta)\,\sigma^\dagger(\theta,\phi)\sigma(\theta,\phi)
    = \int d\alpha\,\tilde\sigma^\dagger(\alpha) \tilde\sigma(\alpha)
    = \eta_\theta\sigma^\dagger\sigma,
\end{equation}
we have for the Poisson process
\begin{equation}
  \dexpct{dN} = \Gamma\,dt
     \int d\alpha\expct{\tilde\sigma^\dagger(\alpha) \tilde\sigma(\alpha)}
     =\eta_\theta\Gamma \expct{\sigma^\dagger\sigma}dt.
\end{equation}
Again, $\alpha$ is a random real number corresponding to the result of
the position measurement for a given spontaneous emission event.
The probability density for $\alpha$ is
\begin{equation}
  \eqnarr{2.2}{
  P(\alpha) \arreq \frac{\expct{\tilde\sigma^\dagger(\alpha) \tilde\sigma(\alpha)}}
      {\eta_\theta\expct{\sigma^\dagger\sigma}}
     \\
   \arreq \frac{1}{\expct{\sigma^\dagger\sigma}}\int dz\,|\psi_\mathrm{e}(z)|^2
     \frac{1}{\sqrt{2\pi}\,\delta\alpha}
      \exp\left[-\,\frac{(z-\alpha)^2}{2(\delta\alpha)^2}\right],
  }
  \label{spontemissalphaprob}
\end{equation}
that is, in the case of a localized atomic wave packet,
a Gaussian probability density with variance $(\delta\alpha)^2$.

\subsection{Adiabatic Approximation}

So far, we have seen how the internal and external dynamics of the
atom are intrinsically linked.  Now we would like to focus on the
external atomic dynamics.
To do so, we will take advantage of the natural separation of time
scales of the dynamics.  The internal dynamics are damped at
the spontaneous emission rate $\Gamma$, which is typically on the
order of $\sim\!\! 10^7\;\mathrm{s}^{-1}$.  The external
dynamics are typically much slower, corresponding to kHz or smaller
oscillation frequencies for typical laser dipole traps.
The adiabatic approximation assumes that the internal dynamics
equilibrate rapidly compared to the external dynamics,
and are thus always in a quasi-equilibrium state with respect to the
external state.

\subsubsection{Internal Quasi-Equilibrium}\label{section:quasiequilibrium}

In treating the internal dynamics, we have noted that the 
atom decays, but not why it was excited in the first place.
A resonant, driving (classical) laser field enters in the form 
\cite{Loudon83}
\begin{equation}
  \HD = \frac{\hbar\Omega}{2}\left(\sigma+\sigma^\dagger\right),
  \label{drivingHam}
\end{equation}
where the \textit{Rabi frequency} $\Omega$ characterizes the 
strength of the laser--atom interaction.
In writing down this interaction, we have implicitly made the
standard unitary transformation to a rotating frame where
$\HA=0$.
We have also assumed the driving field propagates along a normal to the
$z$-axis, so we have not written any spatial dependence of the field
in $\HD$.

The usual unconditioned master equation with this interaction, but neglecting
the external motion (that is equivalent to the usual, on-resonance 
optical Bloch equations)
is
\begin{equation}
  \partial_t\rho = -\,\frac{i}{\hbar}[\HD,\rho] + 
    \Gamma\mathcal{D}[\sigma]\rho.
\end{equation}
This equation implies that the expectation value of an operator $A$ 
evolves as
\begin{equation}
  \partial_t\expct{A} = -\,\frac{i}{\hbar}\expct{[A,\HD]} + 
    \Gamma\expct{\sigma^\dagger A\sigma -\frac{1}{2}[\sigma^\dagger\sigma,A]_+}.
\end{equation}
This gives the following equations of motion for the density-matrix elements:
\begin{equation}
  \eqnarr{1.5}{
    \partial_t \rho_\mathrm{ee} \arreq
      \partial_t \expct{\sigma^\dagger\sigma}
      =\frac{i\Omega}{2}\left(\expct{\sigma}-\expct{\sigma^\dagger}\right)
       -\Gamma\expct{\sigma^\dagger\sigma} , \\
    \partial_t \rho_\mathrm{eg} \arreq
      \partial_t \expct{\sigma}
      =\frac{i\Omega}{2}\left(\expct{\sigma^\dagger\sigma}-\expct{\sigma\sigma^\dagger}\right)
       -\frac{\Gamma}{2}\expct{\sigma}.
  }
  \label{OBEsinsigmaform}
\end{equation}
The remaining matrix elements are determined by $\rho_\mathrm{ge}=\rho_\mathrm{eg}^*$
and $\rho_\mathrm{gg} = \expct{\sigma\sigma^\dagger}= 1-\,\expct{\sigma^\dagger\sigma}$.
Setting the time derivatives to zero, we can solve these equations
to obtain
\begin{equation}
  \eqnarr{1.5}{
     \expct{\sigma^\dagger\sigma}\ss \arreq \frac{\Omega^2/\Gamma^2}{1+2\Omega^2/\Gamma^2} , \\
     \expct{\sigma}\ss \arreq \frac{-i\Omega/\Gamma}{1+2\Omega^2/\Gamma^2} ,
  }
\end{equation}
for the internal steady-state of the atom.

\subsubsection{External Master Equation}

To make the adiabatic approximation and eliminate the internal dynamics,
we note that there is no effect on the external dynamics apart
from the slow center-of-mass motion in the potential $V(x)$ and
the collapses due to the detection events.
When the internal timescales damp much more quickly than the external
time scales, we can make the replacement
\begin{equation}
  \expct{\sigma^\dagger\sigma}\longrightarrow
  \expct{\sigma^\dagger\sigma}\ss
\end{equation}
in the master equation (\ref{partialefficiencyjumpmasteqalpha}).
Also, in steady state, the internal equations of motion
(\ref{OBEsinsigmaform})
give
\begin{equation}
  \expct{\sigma^\dagger\sigma} 
   = \frac{\Omega^2/\Gamma^2}{1+\Omega^2/\Gamma^2}
  \expct{\sigma\sigma^\dagger},
\end{equation}
so that the ground- and excited-state populations are proportional.
When we also account for the atomic spatial dependence,
this argument applies at each position $z$, so that we can
write
\begin{equation}
  |\psi_\mathrm{e}(z)|^2
   = \frac{\Omega^2/\Gamma^2}{1+\Omega^2/\Gamma^2}
  |\psi_\mathrm{g}(z)|^2,
\end{equation}
where we are using the general decomposition 
\begin{equation}
  \langle z|\psi\rangle = \psi_\mathrm{e}(z)\ket{\mathrm{e}} + 
    \psi_\mathrm{g}(z)\ket{\mathrm{g}}
  \label{egequiv}
\end{equation}
for the atomic state vector.
Thus, the spatial profile of the atom is independent of its internal state,
so we need not assign multiple wave functions
$\psi_\mathrm{g}(z)$ and $\psi_\mathrm{e}(z)$ 
to different internal states of the atom.

Furthermore, we will take a partial trace over the internal degrees of freedom by defining
the external density operator
\begin{equation}
  \rhoext := 
    \expct{\mathrm{e}|\rho|\mathrm{e}}+
    \expct{\mathrm{g}|\rho|\mathrm{g}}.
\end{equation}
The result of applying the same partial trace on the master equation is
\begin{equation}
  \eqnarr{2.1}{
  d\rhoext
    \arreq-\,\frac{i}{\hbar}[\HCM,\rhoext]dt \\&&\ds
     {} +\gamma\int d\Omega\,[1-T(\theta)]\, f(\theta,\phi)\,\mathcal{D}[e^{-i\kL z\cos\theta}]\rhoext\,dt \\&&\ds
  {}+\left[ \frac{A(\alpha)\,\rhoext\,A^\dagger(\alpha)}{\expct{A^\dagger(\alpha)A(\alpha)}}-\rhoext\right]dN,
  }
  \label{partialefficiencyjumpmasteqext}
\end{equation}
where
\begin{equation}
  \eqnarr{1.5}{
      \dexpct{dN}\arreq \eta_\theta\gamma\,dt\\
      \tilde \sigma(\alpha)&{}=:{}&\sigma A(\alpha)\\
      \gamma &{}:={}&\Gamma\expct{\sigma^\dagger\sigma}.
  }
\end{equation}
The form (\ref{partialefficiencyjumpmasteqext}) follows from the
fact that the density operator $\rho$
factorizes into external and internal parts,
as we saw in Eq.~(\ref{egequiv}).
Also, Eq.~(\ref{spontemissalphaprob}) becomes
\begin{equation}
  \eqnarr{2.2}{
  P(\alpha) 
   \arreq \int dz\,|\psi(z)|^2
     \frac{1}{\sqrt{2\pi}\,\delta\alpha}
      \exp\left[-\,\frac{(z-\alpha)^2}{2(\delta\alpha)^2}\right],
  }
  \label{spontemissalphaprobadaib}
\end{equation}
where $\psi(z)$ is the effective state-independent wave function for the 
atom.
When the external state is not pure, we simply make the substitution
$|\psi(z)|^2\longrightarrow \bra{z}\rho_\mathrm{ext}\ket{z}$
in Eq.~(\ref{spontemissalphaprobadaib})
to handle this.

Now we have what we want: a master equation for the atomic center-of-mass
state that exhibits localizing collapses due to a physical measurement
process.  What we essentially have is continuous evolution, with
the end of each interval of mean length $(\eta_\theta\gamma)^{-1}$
punctuated by a POVM-type reduction of the form
$\rho\longrightarrow A(\alpha)\rho A^\dagger(\alpha)$.
But note that here there is extra disturbance for the amount of information
we gain, because the aperture only picks up a fraction of the
available information.  We will return to this point shortly.

\subsection{White-Noise Limit}

We now have a POVM with a form similar to Eq.~(\ref{xpovm}), but 
we still have a quantum-jump master equation for a position measurement
that does not look like Eq.~(\ref{SME}).
However, we can note that the Gaussian form of the collapse operator
$A(\alpha)$ is applied to the state after every time interval
of average length $\Delta t=(\eta_\theta\gamma)^{-1}$.  In the regime of 
slow atomic center-of-mass motion, the collapses come quickly
compared to the motion.  Then it is a good approximation to 
take the formal limit $\Delta t\longrightarrow 0$, while keeping
the rate of information gain constant.
(Note that the same result arises in homodyne detection, 
where the emitted light interferes with a strong phase-reference field,
without any coarse-graining approximation.)

\subsubsection{Quantum-State Diffusion}

Comparing Eq.~(\ref{spontemissalphaprobadaib}) with Eq.~(\ref{probalphaPOVM}),
we see that they are the same if we identify
\begin{equation}
  4k\Delta t = \frac{1}{2(\delta\alpha)^2}.
\end{equation}
Note that $k$ here refers to the \textit{measurement strength}, not the
wave number $\kL$ of the scattered light.
Solving for the measurement strength,
\begin{equation}
  k = \frac{\eta_\theta\gamma}{8(\delta\alpha)^2}
    = \frac{\pi^2\eta_\theta\gamma(\delta\theta)^2}{2\lambda^2}.
\end{equation}
Repeating the procedure of Section~\ref{section:continuousmeasurement},
we can take the limit $\Delta t\longrightarrow 0$ with 
$k$ fixed.
The resulting master equation, in ``quantum-state diffusion'' form, is
\begin{equation}
  \eqnarr{2.1}{
  d\rhoext
    \arreq-\,\frac{i}{\hbar}[\HCM,\rhoext]dt \\&&\ds
     {} +\gamma\int d\Omega\,[1-T(\theta)]\, f(\theta,\phi)\,\mathcal{D}[e^{-i\kL z\cos\theta}]\rhoext\,dt \\&&\ds
     {} + 2k\mathcal{D}[z]\rhoext\,dt
     {} + \sqrt{2\eta_\phi k}\mathcal{H}[z]\rhoext\,dW.
  }
  \label{spontemissQSD}
\end{equation}
The form here is the same as in
Eq.~(\ref{SME}), except for an extra ``disturbance term'' representing
the undetected photons.
We have also added an extra efficiency $\eta_\phi$ to model aperturing in
the $\phi$ direction and other effects such as
the intrinsic (quantum) efficiency of the imaging detector.

\subsubsection{Diffusion Rates}

To simplify the master equation (\ref{spontemissQSD}), we will analyze the
diffusion rates due to the second and third terms (proportional
to $\gamma$ and $k$, respectively). From the analysis of 
Eqs.~(\ref{xmeasmoments}), recall that the
term $2k\mathcal{D}[z]\rhoext\,dt$ causes diffusion in momentum
at the rate
\begin{equation}
  D_k = 2\hbar^2 k=\frac{\eta_\theta}{4}\gamma\hbar^2\kL^2(\delta\theta)^2.
\end{equation}
This is the disturbance corresponding to the information gain.
The relation $k=D_k/(2\hbar^2)$ will be useful below.

We can compute the \textit{total} diffusion rate due to the spontaneously
emitted photons as follows.
Each photon emission causes a momentum kick of magnitude
$\hbar\kL\cos\theta$, and the spontaneous emission rate is $\gamma$.
Averaging over the angular photon distribution,
the diffusion rate becomes
\begin{equation}
  \DSE = \gamma\hbar^2\kL^2\int d\Omega\, f(\theta,\phi)\cos^2\theta =
     \frac{\gamma\hbar^2\kL^2}{5}.
\end{equation}
On the other hand, the diffusion rate due only to the detected photons
is
\begin{equation}
  \eqnarr{1.8}{
    D_\theta\arreq\gamma\hbar^2\kL^2 \int d\Omega\, T(\theta)\,f(\theta,\phi)
      \cos^2\theta\\
     \arreq \gamma\hbar^2\kL^2\frac{3}{4}\int_0^\pi d\theta\,
        \sin^3\theta\,\cos^2\theta \exp\left[-\,\frac{2(\theta-\pi/2)^2}{(\delta\theta)^2}\right]\\
     &{}\approx{}&\ds
     \frac{\eta_\theta}{4}\gamma\hbar^2\kL^2(\delta\theta)^2,
  }
\end{equation}
where we used the fact that $\delta\theta$ is small.  This is
precisely the same rate as $D_k$, since they are two different
representations of the same physical process.

We see now that the second and third terms of Eq.~(\ref{spontemissQSD}) 
have the same effect
of momentum diffusion, but at different rates.
We can formally combine them to obtain
\begin{equation}
  \eqnarr{2.1}{
  d\rhoext
    \arreq-\,\frac{i}{\hbar}[\HCM,\rhoext]dt \\&&\ds
     {} + 2\keff\mathcal{D}[z]\rhoext\,dt
     {} + \sqrt{2\etaeff\keff}\mathcal{H}[z]\rhoext\,dW,
  }
  \label{spontemissQSDeff}
\end{equation}
where the effective measurement strength is
\begin{equation}
  \keff = \frac{\DSE}{2\hbar^2} = \frac{\gamma\kL^2}{10},
\end{equation}
and the effective measurement efficiency is
\begin{equation}
  \etaeff = \frac{\eta_\phi k}{\keff}
     = \frac{5}{4}\eta_\phi\eta_\theta(\delta\theta)^2.
\end{equation}
Notice that since $\delta\theta$ is assumed small, 
the \textit{apparent} efficiency $\etaeff$
derived from comparing the information rate to the disturbance rate,
is much smaller than the photon-detection efficiency
of $\eta_\phi\eta_\theta$.  Evidently, the photons radiated
near $\theta=\pi/2$ are much less effective compared to the 
photons radiated near $\theta=0$ or $\pi$.
This result is counterintuitive when considering typical imaging setups
as we have considered here, but suggests that other ways of 
processing the radiated photons (e.g., measuring the phase of
photons radiated closer to the $z$-axis) are more effective than camera-like
imaging. 

\section{Conclusion}

We have presented what we hope is a readily accessible introduction to
continuous measurements in quantum systems.  If you have read and
digested most of the above, you should have a good basic understanding
of how to treat such measurements and manipulate the equations that
describe them.  There is now a considerable literature discussing such
measurements in a variety of systems, and here we give a brief
overview of this literature so as to provide a pointer to further
reading.  We have already mentioned that continuous measurement has
many applications in areas such as feedback control and metrology, and
references on these topics have been given in the introduction.  The
early pioneering work on continuous measurement may be found in
Refs.~\cite{Srinivas81,Barchielli82, Gisin84, Barchielli85, Diosi86,
BelavkinLQG, Diosi88, Belavkin89}.  Derivations of continuous
measurements driven by Gaussian noise in quantum-optical systems are
given in Refs.~\cite{Carmichael89, Wiseman93, DJ}, and further
applications in quantum optics may be found in Refs.~\cite{Wiseman93c,
Carm93, Garraway94, Wiseman95, Wiseman96, Plenio98, WisemanLinQ}.
Derivations and applications of stochastic Schr\"{o}dinger equations
with jump (Poisson) processes---developed originally in quantum optics
as a tool for the simulation of master equations using the ``Monte
Carlo'' method, as in Section~\ref{section:photodetection}---may be
found in~\cite{Zoller87, Hegerfeldt91, Holland91, Dalibard92,
Gardiner92, PlenioPhD, Hegerfeldt96, Garraway94, Holland96, Plenio98}.
A treatment of continuous measurement in a solid-state system is
given in~\cite{Korotkov}, and further applications in these systems
may be found in~\cite{Goan01, Hopkins03, Brun03, Santamore04,
Ruskov05, Sarovar05, Jordan06}.  Last, but not least, if the reader is
interested in treatments of quantum continuous measurements using the
rigorous mathematical language of filtering theory, these may be found
in Refs.~\cite{Bouten06, Belavkin93, Belavkin94, BelavkinLQG}.  Other
rigorous treatments are given in Refs.~\cite{Barchielli85,
Barchielli93}.

\section{Acknowledgments}

The authors would like to thank Tanmoy Bhattacharya, Michael Raymer,
Elizabeth Schoene,
and Jeremy Thorn for insightful comments and corrections.
D.A.S. acknowledges support from the National Science Foundation, and 
K.J. acknowledges support from the Hearne Institute for Theoretical 
Physics, the National Security Agency, the Army
Research Office, and the Disruptive Technologies Office. 


\onecolumngrid
\vspace{5mm}
\twocolumngrid  

\noindent
\textbf{Kurt Jacobs} is an Assistant Professor of Physics at the University of
Massachusetts at Boston.  He completed a master's degree in physics
with Dan Walls at Auckland University in 1995, and a Ph.D. with Peter
Knight at Imperial College, London, in 1998.  During his time at
Auckland he was introduced to classical continuous measurement theory
by Sze Tan, and was exposed to the recently developed field of quantum
continuous measurement particularly in interactions with Howard
Wiseman who was then a postdoc at Auckland.  He continued to work on
this subject during his Ph.D., and then joined Los Alamos National
Laboratory where he worked with Salman Habib on problems in quantum
feedback control and the quantum-to-classical transition.  His work
currently focuses on these two areas, including applications of
continuous measurement and control in atom optics and
quantum-nano-electro mechanics.

\noindent
\textbf{Daniel Steck} is an Assistant Professor of Physics at the University of
Oregon.  He performed his Ph.D. work, involving experiments in quantum
chaos and transport, with Mark Raizen at the University of Texas at
Austin.  He was then a postdoctoral fellow at Los Alamos National
Laboratory with Salman Habib, working on the theory of quantum
measurement and feedback control.  His work now focuses on the study
of continuous quantum measurements and the transition between quantum
and classical dynamics in ultracold-atom experiments.

\end{document}